\newcommand \Mpc {h^{-1}{\rm Mpc}}
\newcommand \kpc {h^{-1}{\rm kpc}}
\newcommand \arcs{\hbox{$^{\prime\prime}$}}
\newcommand \arcm{\hbox{$^{\prime}$}}

\newcommand \farcm{\hbox{$.\!\!^{\prime}$}}

\newcommand \kms {{\rm km~s}^{-1}}
\newcommand \msun {h^{-1} M_\odot}
\newcommand \beqn {\begin{equation}}
\newcommand \eeqn {\end{equation}}
\newcommand \ncz {1323 }
\newcommand \nczmohr {246 } 
\newcommand \nczmahdavi {84 } 
\newcommand \nczuzc {158 } 
\newcommand \nczother {15} 
\newcommand \nczcairns {820 } 
\newcommand \ncznew {1092 } 
\documentclass[12pt,preprint]{aastex}
\usepackage{epsfig}
\usepackage{natbib}

\begin{document}
\bibliographystyle{apj}

\title{Mass Profile of the Infall Region of the Abell 2199 Supercluster}

\author{K. Rines\altaffilmark{1}, M.J. Geller\altaffilmark{2},
A. Diaferio\altaffilmark{3}, A. Mahdavi\altaffilmark{4,5}, 
J.J. Mohr\altaffilmark{6}, and G. Wegner\altaffilmark{7}}
\email{krines@cfa.harvard.edu}

\altaffiltext{1}{Harvard-Smithsonian Center for Astrophysics, 60 Garden St,
Cambridge, MA 02138 ; krines@cfa.harvard.edu}
\altaffiltext{2}{Smithsonian Astrophysical Observatory; mgeller@cfa.harvard.edu}
\altaffiltext{3}{Universit\`a degli Studi di Torino,
Dipartimento di Fisica Generale ``Amedeo Avogadro'', Torino, Italy; diaferio@ph.unito.it}
\altaffiltext{4}{Instiitute for Astronomy, University of Hawaii, 2680
Woodlawn Drive, Honolulu, HI 96822}
\altaffiltext{5}{Chandra Fellow}
\altaffiltext{6}{Departments of Astronomy and Physics, University of 
Illinois, 1002 W. Green St. Urbana, Il  61801; jmohr@uiuc.edu}
\altaffiltext{7}{Department of Physics and Astronomy, Dartmouth College, Hanover, NH 03755; gaw@bellz.dartmouth.edu}

\begin{abstract}

Using a redshift survey of \ncz galaxies (\ncznew new or remeasured)
in a region of 95 square degrees centered on the nearby galaxy cluster
Abell 2199, we analyze the supercluster containing A2199, A2197, and
an X-ray group.  The caustic technique accurately reproduces the true
mass profiles of simulated simple superclusters (i.e., superclusters
where the virial mass of one cluster is 2-10 times the virial mass of
all other clusters in the supercluster).  We calculate the masses of
the two main components of A2197 (A2197W and A2197E) using archival
X-ray observations and demonstrate that the A2199 supercluster is
simple (the mass of A2199 is 5 and 12 times larger than A2197W
and A2197E respectively) and thus that the caustic technique should
yield an accurate mass profile.  The masses of A2199, A2197W, and
A2197E within $r_{500}$ (the radius within which the enclosed density
is 500 times the critical density) are 22.0, 3.8, and 1.7$\times
10^{13}\msun$.  The  mass profile is uncertain by $\sim$30\%
within $3~\Mpc$ and by a factor of two within $8~\Mpc$ and is one of
only a few for a supercluster on such large scales.  Independent X-ray
mass estimates agree with our results at all radii where they overlap.
The mass profile strongly disagrees with an isothermal sphere profile
but agrees with profiles suggested by simulations.  We discuss the
interplay of the supercluster dynamics and the dynamics of the bound
subclusters.  The agreement between the infall mass profile and other
techniques shows that the caustic technique is surprisingly robust for
simple superclusters.

\end{abstract}


\section{Introduction}

Clusters of galaxies are important probes of the distribution of
matter on intermediate scales.  Clusters are surrounded by infall
regions where the galaxies are falling into the gravitational
potential well of the cluster, but have not yet reached
equilibrium. Many, perhaps most, of the galaxies in this region are on
their first pass through the cluster. They populate a regime between that
of relaxed cluster cores and the surrounding large-scale structure
where the transition from linear to non-linear clustering occurs.
Recent papers have explored infall regions using two-body dynamics of
binary clusters \citep[e.g., ][]{mw97}, the virial theorem in
superclusters \citep{small98}, and weak lensing \citep{kaiserxx}.
\citet{ellingson01} demonstrate that these infall regions are the
site of significant galaxy evolution and are key to understanding the
Butcher-Oemler \citep{1984ApJ...285..426B} effect.  Two recent papers
\citep{rqcm, rines01b} demonstrate that the infall regions of clusters
(the Shapley Supercluster and Abell 2199 respectively) provide
evidence of hierarchical structure formation on supercluster scales.

Because clusters are not in equilibrium outside the virial radius,
neither X-ray observations nor virial analysis provide accurate mass
determinations at large radii. There are now two methods of
approaching this problem: weak gravitational lensing \citep{kaiserxx}
and kinematics of the infall region \citep[][hereafter DG and D99
respectively]{dg97,diaferio1999}.  In redshift space, the infall
regions of clusters form a characteristic trumpet-shaped pattern.
These caustics arise because galaxies fall into the cluster as the
cluster potential overwhelms the Hubble flow
\citep{kais87,rg89}. Under simple spherical infall, the
galaxy phase space density becomes infinite at the caustics. DG
analyzed the dynamics of infall regions with numerical simulations and
found that in the outskirts of clusters, random motions due to
substructure and non-radial motions make a substantial contribution to
the amplitude of the caustics which delineate the infall regions
\citep[see also][]{vh98}. DG showed that the amplitude of the caustics
is a measure of the escape velocity from the cluster; identification
of the caustics therefore allows a determination of the mass profile
of the cluster beyond the virial radius. DG find that nonparametric
measurements of caustics in simulated clusters yield mass
profiles accurate to $\sim$50\% on  scales up to $10~\Mpc$.

The caustic technique places no requirements on the equilibrium state
of the cluster; rather, it assumes only that galaxies trace the
velocity field. Indeed, simulations suggest that little or no velocity
bias exists on linear and mildly non-linear scales
\citep{kauffmann1999a,kauffmann1999b}.  \citet{vh98} used  
simulations to explore an alternative parametric analysis of the
infall region using a maximum likelihood method. Their technique
requires assumptions about the functional forms of the density profile
and the velocity dispersion profile.  \citet[][hereafter GDK]{gdk99}
applied the kinematic method of DG to the infall region of the Coma
cluster. They successfully reproduced the X-ray derived mass profile
and extended direct determinations of the mass profile to
$\sim$10$~\Mpc$.  This method has also been applied to the Shapley 
Supercluster \citep{rqcm}, A576 \citep{rines2000}, A1644
\citep{tustin} and the Fornax cluster \citep{drink}.

We apply the method of DG and D99 to A2199, one of the
richest, most regular nearby clusters of galaxies \citep[][hereafter
MVFS]{mvfs}.  \citet{rines01b} presented a preliminary view of this
system and demonstrated that it contains several infalling, bound
subclusters.  This system provides an important test of the robustness
of the caustic technique in simple superclusters (i.e., those
dominated by a cluster 2-10 times more massive than any other system
in the supercluster).

We describe numerical simulations of simple superclusters in $\S$ 2 to
demonstrate the validity of the caustic technique for these
systems. We describe the spectroscopic observations in $\S$ 3.  In
$\S$ 4, we show that A2197 is a superposition of groups and
demonstrate that A2199/A2197 is indeed a simple supercluster.  We
determine the amplitude of the caustics, calculate the mass profile,
and compare the infall mass profile to other estimators in $\S 5$. We
discuss our results in $\S 6$ and conclude in $\S 7$.  The physical
scale at the redshift of the supercluster ($cz=9156~\kms$) is $1^\circ
= 1.54~\Mpc$ ($H_0 = 100~h~\kms, \Omega _m = 0.3, \Omega _\Lambda =
0.7$).

\section{Simulations of Simple Superclusters}

The caustic technique as developed by D99 is successful in reproducing
the mass profiles of isolated clusters.  Before applying the technique
to more complex systems, we need to verify that this technique still
produces accurate mass profiles in the presence of significant
substructure in the infall region.  We extracted three simple superclusters from the $\Lambda$CDM cosmological
simulation of the GIF collaboration, where galaxy formation and
evolution within dark matter halos are modelled with a semi-analytic
technique \citep{kauffmann1999a,kauffmann1999b}.  Here, we
define simple superclusters as those dominated by a cluster 2-10 times
more massive than any other systems in the supercluster.  These three
systems have mass, $M_1$, within $r_{200}$ (roughly equivalent to the
virial mass) for the central system greater than $10^{14}\msun$; the
next largest subhalo has a mass $M_2$ between $10^{13}\msun$ and
$10^{14}\msun$ and lies outside the virial radius of the main cluster.
Table \ref{sims} lists the properties of the simulated
superclusters. Column [1] gives the system ID, Column [2] is the mass
$M_1$ of the largest system in the supercluster, Column [3] is the
mass ratio $M_1/M_2$, Column [4] and Column [5] list the mass and 3-D
distance of the four subhalos (with $M>10^{13}\msun$) closest to the
largest system.  The mass ratio $M_1/M_2$ ranges between 2 and 6 to
investigate the sensitivity of the mass estimator to companions of
various relative sizes.  We restrict our analysis to $\Lambda$CDM
simulations because these more accurately reproduce the appearance of
the observed caustics in nearby systems (see references in $\S 1$);
caustics in $\tau$CDM simulations are less well-defined.  D99 suggests
that the degree of delineation of the caustics is a cosmological
indicator; caustics are better defined (have fewer interlopers) in
low-$\Omega_m$ universes because the accretion rate in the present
epoch decreases with decreasing $\Omega_m$.

\begin{figure}
\figurenum{1}
\plotone{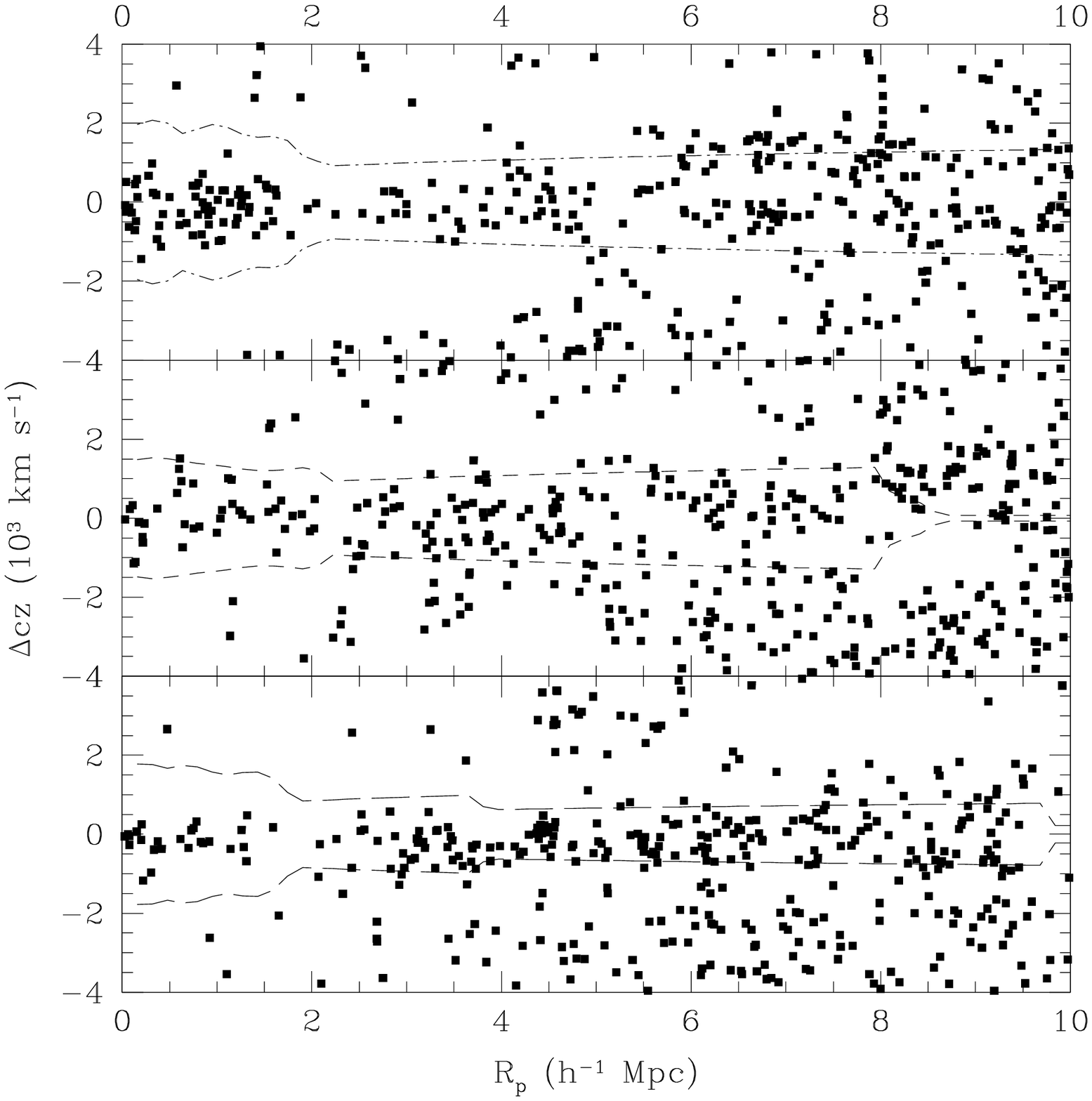} 
\caption{\label{causpair1} Redshift versus radius and the caustics for
a simulated supercluster (ID 121) with mass ratio $M_1/M_2=2.0$ (see
text for definition) projected along three different lines of
sight. We use different line styles to enable comparison with the
estimated mass profiles in Figures \ref{masspair} and \ref{massdelchi}.} 
\end{figure}

\begin{figure}
\figurenum{2}
\plotone{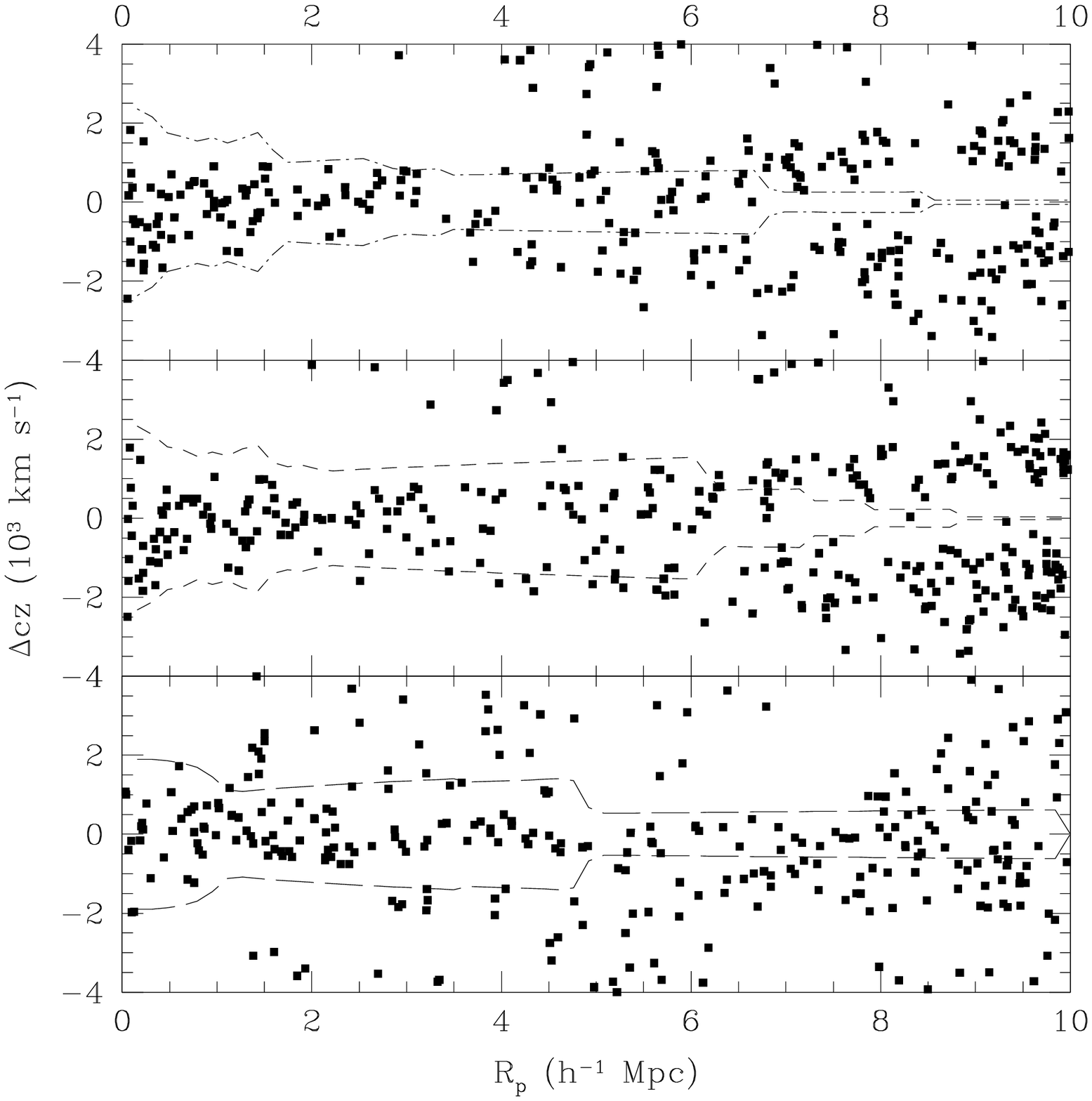} 
\caption{\label{causpair2} Same as Figure \ref{causpair1} for the
simulated supercluster ID 156 with mass ratio $M_1/M_2=4.4$. } 
\end{figure}

\begin{figure}
\figurenum{3}
\plotone{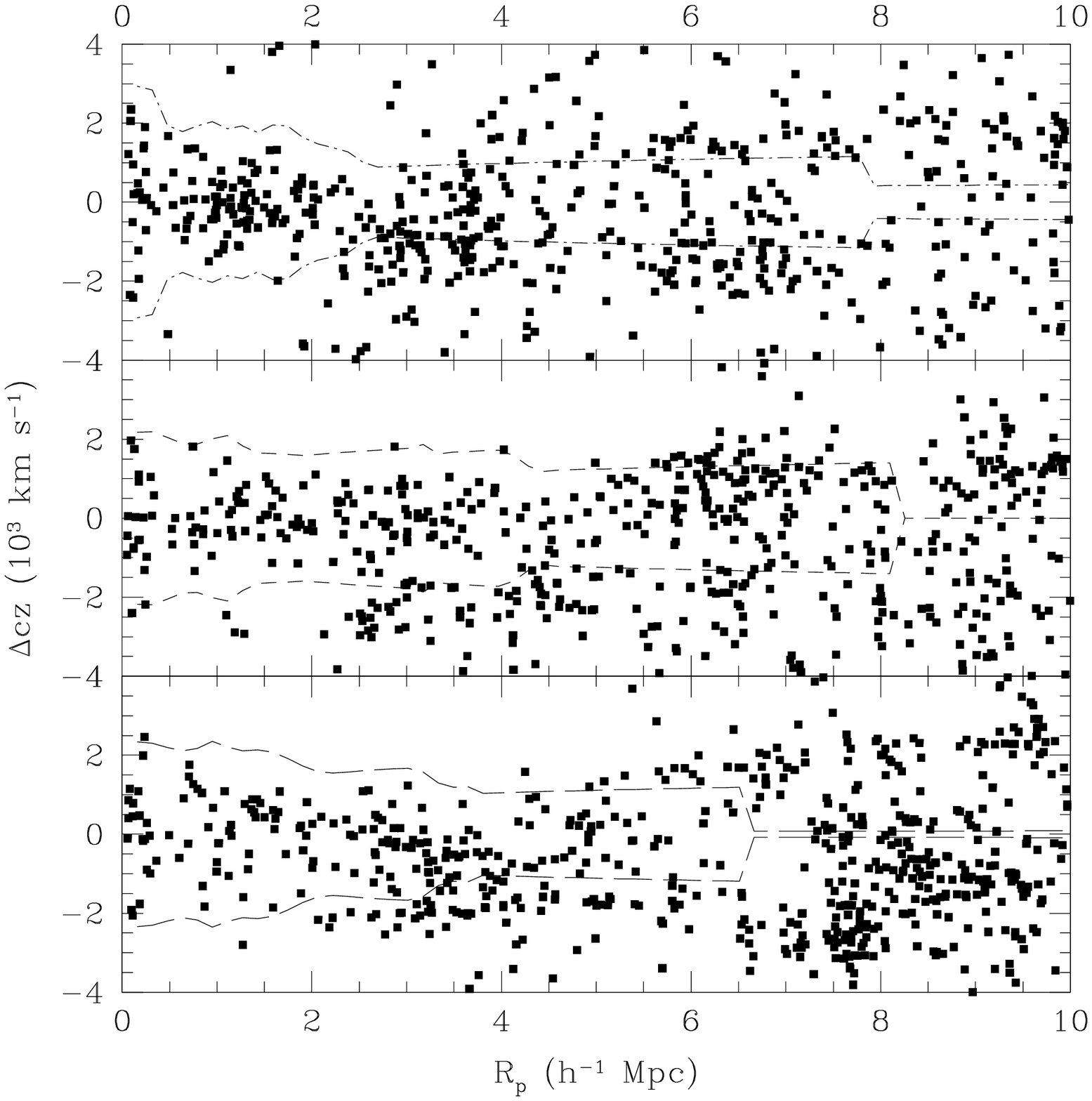} 
\caption{\label{causpair3} Same as Figure \ref{causpair1} for the
simulated supercluster ID 162 with mass ratio $M_1/M_2=6.2$. } 
\end{figure}

\begin{figure}
\figurenum{4}
\plotone{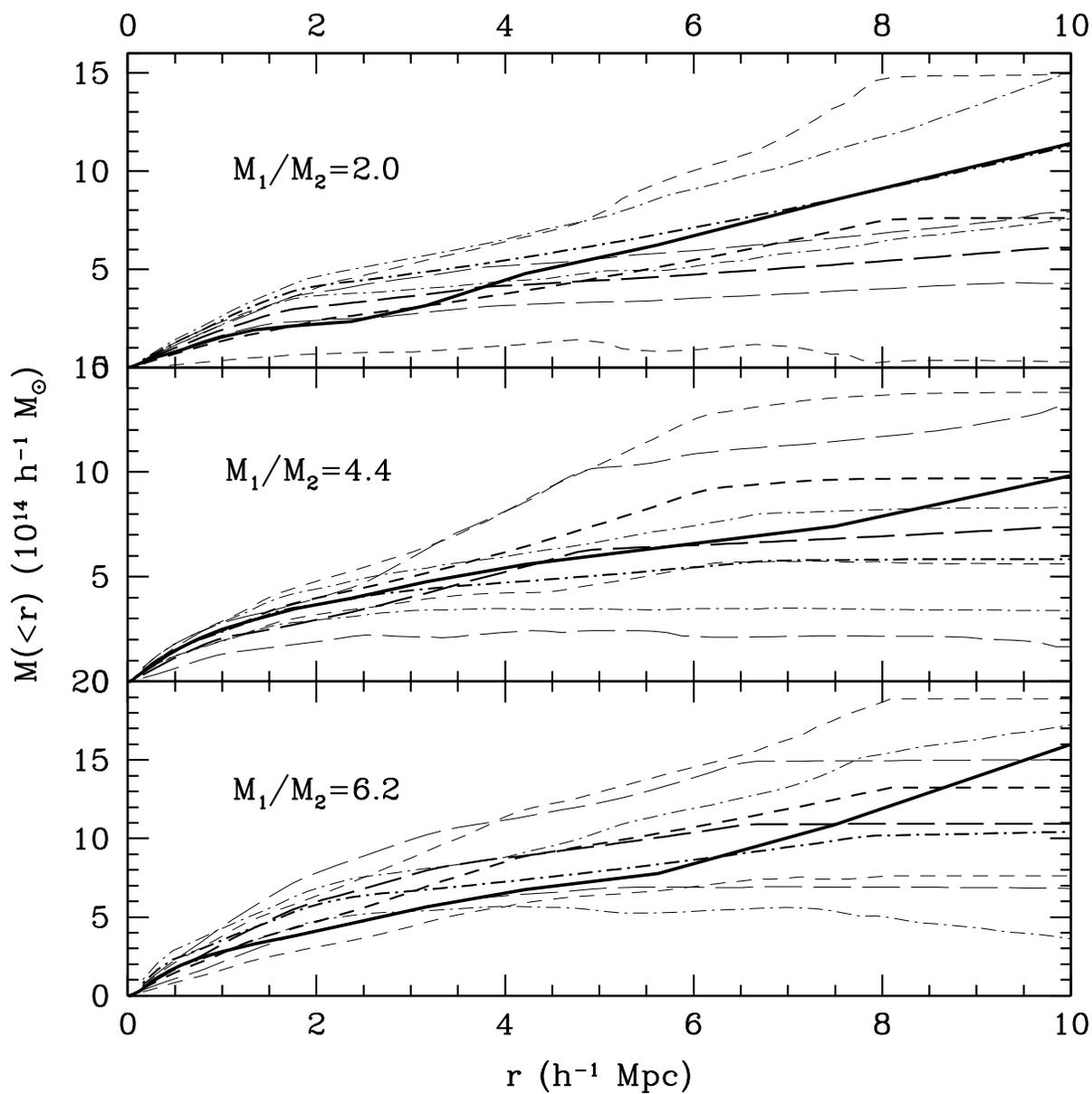}
\caption{\label{masspair} Comparison of caustic mass profiles to
true mass profiles for simulated superclusters. The three sets of
lines show the 1-$\sigma$ ranges of the infall mass profile from three
different lines of sight.  The line styles are the same as in Figures
\ref{causpair1}-\ref{causpair3}.} 
\end{figure}

\begin{figure}
\figurenum{5}
\plotone{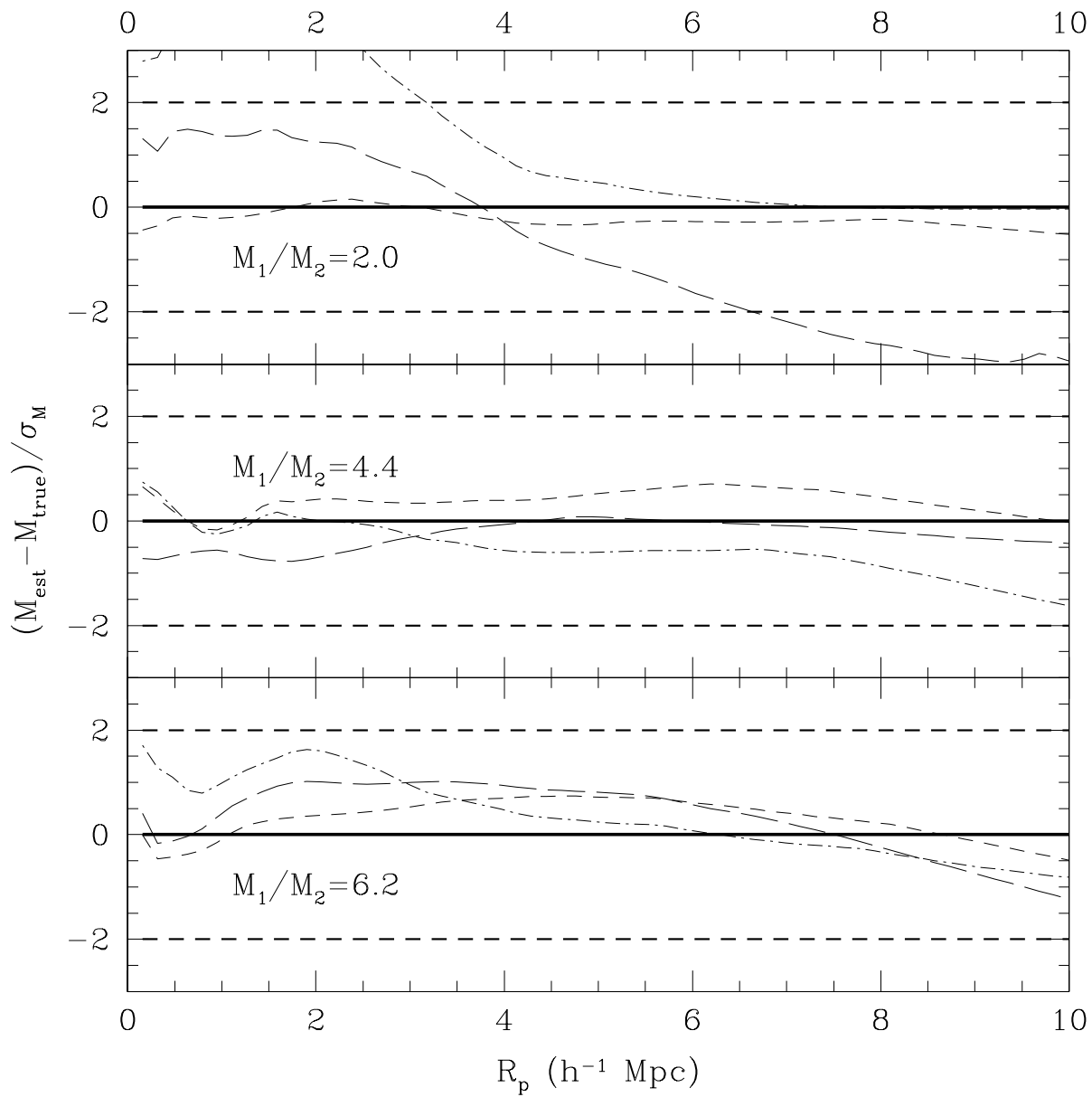}
\caption{\label{massdelchi} Difference between estimated and true
masses of the simulated superclusters in units of the uncertainty in
the estimated mass. The estimates differ by more than 2$\sigma$ from
the true mass profile only in the supercluster with $M_1/M_2=2.0$. The
line styles are the same as in Figures
\ref{causpair1}-\ref{causpair3}.} 
\end{figure}

Figures \ref{causpair1}-\ref{causpair3} display the caustics
(calculated according to the prescription of D99) for these systems as
viewed from three different lines of sight. Redshift diagrams include
all galaxies brighter than $M_B=-18.5+5\mbox{log}h$.  As in isolated clusters
(D99), some lines of sight are more favorable than others.  Figure
\ref{masspair} compares the mass profiles inferred from the caustics
to the true radial mass profiles.  Figure \ref{massdelchi} shows the
difference between the estimated and true mass profiles in units of
the uncertainty in the estimated mass.  The estimated mass differs by
more than 2-$\sigma$ only for some lines of sight to the supercluster
with $M_1/M_2=2.0$.  The bias in the mass estimates appears to be
small, although the small number of systems considered prevents firm
conclusions.  Thus, the caustic method is robust to the presence of
massive subhalos provided the mass ratio (measured within $r_{200}$)
is greater than about 2.0.  We show in $\S 4$ that A2197 (actually
A2197W$+$A2197E) is much less massive than A2199, the caustic
technique should thus yield an accurate mass profile for the A2199
supercluster.

\subsection{Model Mass Profiles}

We now investigate whether the mass profiles of the simulated simple
superclusters are well described by simple analytic models.  The
simplest model of a self-gravitating system is a singular isothermal
sphere (SIS). The mass of the SIS increases linearly with radius.
\citet{nfw97} and \citet{hernquist1990} propose two-parameter models
based on CDM simulations of haloes.  At large radii, the NFW mass
profile increases as ln$(r)$ and the mass of the Hernquist model
converges.  The NFW mass profile is 
\beqn
M(<r) = \frac{M(a)}{\mbox{ln}(2) - \frac{1}{2}}[\mbox{ln}(1+\frac{r}{a})-\frac{r}{a+r}]
\eeqn
where $a$ is the scale radius and $M(a)$ is the mass within $a$. We fit
the parameter $M(a)$ rather than the characteristic density $\delta_c$
($M(a) = 4\pi \delta_c \rho_c a^3 [\mbox{ln}(2) - \frac{1}{2}]$ where
$\rho_c$ is the critical density) because $M(a)$ and $a$ are much less
correlated than $\delta_c$ and $a$.  The Hernquist mass profile is
\beqn
M(<r) = M \frac{r^2}{(r+a)^2}
\eeqn
where $a$ is the scale radius and $M$ is the total mass. Note that
$M(a) = M/4$. The SIS mass profile is 
\beqn
M(<r) = M(a=0.5~\Mpc) \frac{r}{0.5~\Mpc}
\eeqn
where we arbitrarily set the scale radius $a=0.5~\Mpc$.  We assume 10\%
uncertainties in the true mass profiles and minimize $\chi ^2$. Table
\ref{simmpfits} lists the best-fit parameters $a$ (fixed for SIS) and
$M(a)$ for the three models for the simulated superclusters.  Because
the individual points in the mass profile are not independent, the
absolute values of $\chi_\nu ^2$ listed in Table \ref{simmpfits} are
indicative only, but it is clear that the NFW and Hernquist profiles
provide acceptable fits to the infall mass profile; the SIS is
excluded for all estimates. The NFW profile generally provides a
better fit to the data than the Hernquist profile. A non-singular
isothermal sphere mass profile yields results similar to the SIS;
thus, we only report our results for the SIS.

\section{Observations}

\subsection{Spectroscopy}

We have collected \ncz redshifts (\ncznew new or remeasured) in a
large region (95 square degrees) surrounding A2199/A2197.  Two of us
(JJM and GW) used the Decaspec \citep{decaspec} at the 2.4-m MDM
telescope on Kitt Peak to obtain \nczmohr redshifts of galaxies in the
central $1^\circ \times 1^\circ$ of this region for a separate Jeans
analysis of the central region of A2199 (Mohr et al.~in preparation).

We used the FAST spectrograph \citep{fast} on the 1.5-m Tillinghast
telescope of the Fred Lawrence Whipple Observatory (FLWO) to measure
\nczcairns spectra of galaxies within $5^\circ$.5
($\approx$8.5$~\Mpc$) of the center of A2199.  FAST is a high
throughput, long slit spectrograph with a thinned, backside
illuminated, antireflection coated CCD detector.  The slit length is
180\arcs; our observations used a slit width of 3\arcs ~and a 300
lines mm$^{-1}$ grating.  This setup yields spectral resolution of 6-8
\AA ~and covers the wavelength range 3600-7200 \AA .  We obtain
redshifts by cross-correlation with spectral templates of
emission-dominated and absorption-dominated galaxy spectra created
from FAST observations \citep{km98}.

We observed infall galaxy candidates in three campaigns. We selected
targets from digitized images of the POSS I. We initially selected
galaxies using the automatic classification system of the Automated
Plate Scanner (APS)$\footnote{The APS databases are supported by the
National Aeronautics and Space Administration and the University of
Minnesota, and are available at http://aps.umn.edu}$; we visually
inspected these targets to eliminate stars. The first campaign yielded
a deep sample in the central $4^\circ \times 4^\circ$ region around
A2199. This sample is complete to 103aE magnitude $E<16.5$ and
consists of 304 redshifts.  The second campaign (379 redshifts) is a
shallower survey ($E<16.1$) of all galaxies within $5.^\circ 5 \approx
8.5~\Mpc$ of A2199.  The third campaign added 137 redshifts for
galaxies which were excluded from the APS catalog for unknown reasons
($\S 3.2$).  The completeness limits are imprecise because the
magnitudes come from multiple plate scans and because we could not
obtain redshifts for some low surface brightness galaxies.  We include
\nczmahdavi redshifts associated with the groups NRGs385 and NRGs388
obtained with FAST for a separate study of the X-ray and optical
properties of groups of galaxies \citep[59 published in][hereafter
MBGR]{mahdavi99,rasscals}.

We collected the remaining redshifts from the Updated Zwicky Catalogue
\citep[][\nczuzc redshifts]{falco99} with a small number (\nczother)
from other sources listed in the NASA/IPAC Extragalactic
Database$\footnote{The NASA/IPAC Extragalactic Database is available
at http://nedwww.ipac.caltech.edu/index.html}$
\citep{koss,strauss,rc3,freud,zab,haynes,hill}. 
Table~\ref{redshifts} contains our redshift data. The table includes
all redshifts obtained with FAST and from the literature as well as
redshifts obtained with Decaspec for all galaxies brighter than
$E$=16.1 (the remaining 188 will appear in Mohr et al.~in
preparation).  This catalog is complete to $E$=16.1 within $5^\circ$.5
of the center of A2199 and includes 38 galaxies outside $5^\circ$.5
but within $6^\circ.5\approx 10~\Mpc$ (Figure \ref{skyplot}, described
in detail in $\S 5$).

\subsection{APS Catalog -- Photometry and Incompleteness}

When matching galaxies from our redshift catalog with a galaxy catalog
from APS, we discovered that several nearby, bright galaxies including
NGC 6160 (a cD galaxy in A2197) are not classified as galaxies by the
APS neural network.  A further search reveals that the APS identified
many of these objects but incorrectly classified them as stars.  The
default magnitude listed for these objects is from the
magnitude-diameter relationship for stars \citep{apsstars}.  Because
galaxies have large diameters, they are assigned bright magnitudes.
We extracted a catalog of all objects classified as stars by APS with
magnitudes $E_{diam}<12$ within $5.^\circ 5$ of A2199.  Visual
inspection of these 4876 objects yielded 160 galaxies (5 of which are
star-galaxy blends), suggesting a misclassification rate of $\sim$3.3\%.
Because these objects include bright, large galaxies like NGC 6160,
the incompleteness in the APS could be significant in nearby galaxy
surveys.

Because of these difficulties, we reanalyzed the POSS I 103aE (red)
plates using SExtractor to calculate object parameters and magnitudes.
We calculated the mean offset between SExtractor MAGBEST magnitudes
and the isophotal magnitudes in the APS catalog to determine the zero
point for each plate.  We next identified galaxy candidates as objects
with MAGBEST$<$16.5, staricity parameter smaller than 0.5, and
ISO7/ISO2 greater than 0.35.  ISO1-7 are the numbers of pixels
included within various isophotes; the latter criterion thus selects
extended objects. We ran SExtractor in CCD mode. We experimented with
running SExtractor in PHOTO mode (appropriate for photographic plates)
after taking the logarithm of the scanned image (this step is needed
to obtain a sufficiently small dynamic range for this mode) and found
nearly identical results for both photometric and shape parameters.

Finally, we visually inspected all galaxy candidates to compile a
complete catalog of galaxies brighter than $E=16.1$.  Our new catalog
includes 970 galaxies, 135 (14\%) of which are not included in the APS
catalog.  These 970 galaxies comprise the majority (71\%) of the
redshift sample; the fainter galaxies in the remainder of the
sample are located mainly near A2199, A2197, and the X-ray groups
NRGs385 and NRGs388.

\subsection{X-Ray Data}

Archival X-ray observations are available for the entire A2199
supercluster from the {\em ROSAT} All-Sky Survey \citep{rass}. Several
longer pointed observations of selected parts of the supercluster are
available in the {\em ROSAT} PSPC and {\em ASCA} archives (Table
\ref{xdata}). \citet{feretti} discuss radio and {\em ROSAT} pointed
observations of the groups NRGs385 and NRGs388. We calculate the
luminosities of the systems in the {\em ROSAT} energy band (Table
\ref{properties}) from the pointed observations assuming that all of
the subsystems are at the distance of A2199 ($cz=$9156$~\kms$, see $\S
5$).  We used the X-ray analysis package ZHTOOLS\footnote{ Available
at http://hea-www.harvard.edu/~alexey/zhtools/} to extract regions
around the sources, subtract background sources, and correct for
vignetting and the exposure maps. We then used
PIMMS\footnote{Available at
http://heasarc.gsfc.nasa.gov/docs/software/tools/pimms.html} to
convert PSPC count rates into fluxes. We assumed Galactic column
density and a Raymond-Smith thermal plasma with an abundance of 0.4
and $kT$= 4.5 keV for A2199 (MVFS), 0.2/1.7 keV for A2197W, and
0.2/1.0 keV for all others (see $\S 5.2$). Figure \ref{skyplot}
displays the RASS data in a large area around A2199 after correcting for
the exposure map and smoothing with a Gaussian with a FWHM of
$2^\prime$.

\section{A2197: An Optically Rich ``Cluster'' Composed of Superposed Groups} 

A2197 has been observed by both {\em ROSAT} and {\em ASCA}.  Figure
\ref{a2197pspc} shows X-ray contours of the 13.5-ksec
combined archival {\em ROSAT} PSPC observation of A2197 (Table
\ref{xdata}) overlaid on an optical image from POSS I. The X-ray
contours clearly show the presence of West-East components separated
by 24$^\prime$ centered on NGC 6160 and NGC 6173 respectively
\citep{muriel1996}.  The galaxy contours (Figure \ref{skyplot}) show
at least two distinct NW-SE components, but the galaxy distribution is
skewed with respect to the X-ray emission \citep[see
also][]{rines01b}.  Because the intragroup gas is expected to better
trace the gravitational potential, we use the X-ray centers in all
subsequent analysis.

We find several distinct X-ray sources in the combined PSPC
observation of A2197.  Table \ref{x97} lists the X-ray luminosities of
sources with obvious optical counterparts.  In addition to A2197W and
A2197E, we identify five galaxies in or near A2197, one background
galaxy, two background Abell clusters (A2187 and A2196), and two
background QSOs.   A2187 is recognized as an X-ray cluster
\citep{bcs,noras} but our determination of the X-ray luminosity of
A2196 is new.  The galaxies have X-ray luminosities of
$L_X(0.1-2.4~\mbox{keV}) = 4-64 \times 10^{40} h^{-2}$ erg s$^{-1}$,
within the range of isolated bright galaxies \citep{andilxsig}.

\subsection{Redshift Distribution}

A2197 shows significant substructure in redshift space. Figure
\ref{v2197} shows a redshift histogram of galaxies within 1.5$~\Mpc$
of A2197W with a bin size of 200~$\kms$.  We calculate $\chi ^2 =
36.3$ for 12 degrees of freedom, indicating that the parent redshift
distribution is non-Gaussian at the 99.9\% confidence level.  A
two-component Gaussian fit yields $\chi ^2 = 10.8$ for 9 degrees of
freedom, an acceptable fit.  The properties of the best-fit components
are $N_1 = 82\pm5$, $\mu_1=8787\pm 33~\kms$, $\sigma_1 = 361\pm
26~\kms$, and $N_2 = 48\pm6$, $\mu_2=9800\pm 35~\kms$, $\sigma_2 =
365\pm 78~\kms$,

We recalculate the mean redshifts and velocity dispersions of the two
subclusters according to the prescription of \citet{danese} but with a
limiting radius of $0.25~\Mpc$, which is small enough that the
groups do not overlap. The mean redshifts of A2197W and A2197E are
$9569\pm151~\kms$ and $8919\pm153~\kms$ based on 11 and 10 members
respectively. The mean redshifts differ by 4.2-$\sigma$. A K-S test
indicates that the redshifts are drawn from different populations at
the 99.6\% confidence level.  Note that the redshifts of NGC 6160 and
NGC 6173 are 9408$~\kms$ and 8842$~\kms$ respectively (Table
\ref{redshifts}), suggesting that they may lie near the centers of the
two groups (Figure \ref{v2197}). 

These tests show that the groups A2197W and A2197E are separated in
redshift space. The two groups overlap each other both on the sky and
in redshift space; they may be interacting in the current epoch.
Apparently, A2197E is a richer group than A2197W but has a lower X-ray
luminosity.  The superposition of the two groups artificially enhances
the apparent optical richness of the system.

\subsection{Surface Brightness Distribution}

We fit a $\beta$ model \citep[]{cff} convolved with the PSPC PSF to
the observed surface brightness of A2197W in the PSPC observation. We
find a core radius $r_c=0\farcm 21 = 11~\kpc$ and $\beta _X = 0.41$.
Note that A2197E is 28\arcm ~off-axis, the PSF of the PSPC at this
radius is $\sim$1\arcm ~and is anisotropic.  Because of these issues,
we do not attempt to fit the surface brightness profile of A2197E.
Our results agree with \citet{muriel1996}.

\subsection{ICM Temperature}

$ASCA$'s broad energy band (0.5-10.0 keV) is particularly useful for
determining cluster temperatures. Because of the poor angular
resolution of $ASCA$, we determine the emission weighted average
temperature within $11'$, or $\sim 0.3~\Mpc$; outside this radius the
cluster emission is overwhelmed by the background. We obtained the
screened data from Goddard Space Flight Center (GSFC). We extract a
spectrum including all photons within a circle of radius $11'$ (44
pixels) centered on the cluster center for GIS data from long
observations of A2197W and A2197E (Table \ref{xdata}). The centroid of
the SIS images agree with the positions of the {\em ROSAT} centroids
within the $\sim 0\farcm 4$ uncertainty in the SIS position
\citep{ascaposn}\footnote{Available at
http://heasarc.gsfc.nasa.gov/docs/asca/newsletters/Contents4.html}.

Using XSPEC (v10.0), we fit the cluster spectrum to a model including
absorption (`wabs') parameterized by the column density of hydrogen
(which we set to the Galactic value) and the standard Raymond-Smith
model \citep{rs77} characterized by temperature, iron abundance,
redshift, and a normalization factor. The iron abundance is measured
relative to cosmic abundance.  We fit the temperature, iron abundance,
and normalization as free parameters.  Because we are interested in
the emission-weighted X-ray temperature, we use only the GIS data
because the SIS field of view does not contain the entire X-ray
emitting regions. 

We use the weighting system developed by \citet{chur}, which properly
accounts for errors in small-number statistics.  Because there are few
counts above 8.0 keV and the calibration uncertainties become large at
low energies, we include only data from $0.8 \to 8.0$ keV.  We fit
the spectrum to slightly different ranges to ensure that the fitted
model parameters are consistent for different choices.

We obtain acceptable fits by assuming that the gas is isothermal.
More complicated models are thus unnecessary. Our best-fit model for
A2197W has an ICM temperature of 1.55$^{+0.16}_{-0.20}$ keV with an
iron abundance 0.13$^{+0.17}_{-0.10}$ cosmic. For A2197E, we find
$kT$=0.96$^{+0.07}_{-0.08}$ keV and an iron abundance
0.20$^{+0.10}_{-0.05}$ cosmic.  Uncertainties are 90\% confidence
limits for one parameter and do not include calibration uncertainties.
The acceptable isothermal model fits are consistent with no large
temperature variations within either A2197W or A2197E, suggesting an
absence of shock fronts which should be present if the groups were in
the process of merging.  Figure \ref{xrayspectrumw}
(\ref{xrayspectrume}) shows the X-ray spectrum and the best-fit model
for A2197W (A2197E). These temperatures are slightly smaller but
consistent with those in \citet{fad} estimated from {\em ASCA} SIS
data for the central regions of the clusters.

Using the best-fit model, we calculate a bolometric flux of $5.1 (4.3)
\times 10^{-12}$ ergs cm$^{-2}$ s$^{-1}$ for A2197W (A2197E).
We calculate a bolometric luminosity of $5.3 (4.5) \times 10^{42}
h^{-2}$ ergs s$^{-1}$.  If we use the {\em ASCA} temperatures and
abundances and the {\em ROSAT} PSPC count rates from the pointed
observations, we estimate bolometric luminosities of $5.0 (4.3) \times
10^{42} h^{-2}$ ergs s$^{-1}$, in good agreement with the {\em ASCA}
estimates.

\subsection{Mass Estimates for A2197W and A2197E}

The mass-temperature relation
\citep{1999ApJ...520...78H,2000ApJ...532..694N, frb2001} gives a
straightforward estimate of the mass of a cluster given its
temperature.  \citet{frb2001} find a relation of $M_{500} = (1.87\pm
0.14) \times 10^{13} T_{keV}^{1.64\pm0.04} \msun$, where $M_{500}$ is
the enclosed mass within a radius $r_{500}$ inside which the mean
density is 500 times the critical density and $T_{keV}$ is the
electron temperature in keV.  Using $T_{keV}$=4.5 (MVFS) for A2199, 
the values of $M_{500}$ for A2199, A2197W, and A2197E are 22.0, 3.8,
and 1.7 respectively (in units of $10^{13} \msun$).  Thus, the mass of
A2199 is $\approx$4 times larger than the combined masses of A2197W and
A2197E.  The A2199 system is therefore similar to the systems
considered in $\S 2$ so we are justified in applying the caustic
technique to A2199.

With the assumptions of hydrostatic equilibrium, negligible
non-thermal pressure, and spherical symmetry, the gravitational mass
inside a radius r is
\begin{equation}
\label{mass1}
M_{tot}(<r) = - \frac{kT}{\mu m_p G} \Bigl({{d\ln \rho_{gas}} \over
{d\ln r}} + {{d\ln T} \over {d\ln r}} \Bigr) r
\end{equation}
\citep{flg80}. For a uniform temperature distribution, the second
term on the right hand side vanishes. We then only need to determine
the gas temperature and the density distribution of the gas to
calculate the gravitational mass.  Under the standard
hydrostatic-isothermal $\beta _x$ model, the mass is related to $\beta
_x$ and $a$ by
\begin{equation}
\label{mass2}
M_{tot}(<r) = \frac{3kT\beta _x r^3}{\mu m_p Ga^2(1 + (\case{r}{a})^2)}
\newline = 5.65\times 10^{13} \beta _x T_{\mbox{keV}} \frac{r^3}{a^2 +
 r^2} \msun
\end{equation}
where $M_{tot}(<r)$ is the total gravitational mass within a radius
$r$ and the numerical approximation is valid for $T_{\mbox{keV}}$ in
keV and $r$ and $a$ in $\Mpc$.

We use Equation~\ref{mass1} to calculate the total mass of A2197W
within 0.5$~\Mpc$; the estimate outside $\sim$0.3$~\Mpc$ is an
extrapolation.  We find a mass of $M \approx 2.2$ and $3.6 \times
10^{13}~\msun$ at $r=0.3$ and $0.5~\Mpc$ respectively.  \citet{wjf97}
use spherical deprojection to estimate $M \approx 3.9 \times
10^{13}~\msun$ at $r=0.28~\Mpc$ for A2197W, in reasonable agreement
with our results.  

We calculate the virial mass of A2197 from galaxies within $1.5~\Mpc$
of the X-ray peak of A2197W.  We find $M_{vir}=5.4\times
10^{14}\msun$, significantly larger than the X-ray mass extrapolated
assuming an isothermal sphere. \citet{girardi98} similarly find that
the virial mass of A2197 is larger than the X-ray mass.  This
discrepancy is easily understood with the recognition that A2197 is
the superposition of at least two groups. Supercluster galaxies not
contained in the groups may further contribute to an artificially
large observed velocity dispersion and hence virial mass.  The
calculated virial mass for A2197 is therefore meaningless.

\section{Defining the Infall Region with Caustics}

Figure \ref{caustics} displays the redshifts of galaxies surrounding
A2199 as a function of projected radius.  The A2199 supercluster is
located within the Great Wall \citep[][ Figure 6b of Falco et
al.~1999]{gh89}, which may complicate the interpretation of the
dynamics of the system.  The expected caustic pattern is easily
visible.  The caustic diagram of A2199 shows fewer interlopers than in
the simulated superclusters, perhaps due to the galaxy formation
recipe used in the simulations or perhaps due to lower infall rates
such as would be present in a cosmology with $\Omega _m < 0.3$.

We perform a hierarchical structure analysis to locate the centroid of
the largest system in the supercluster (D99).  This analysis yields a
position for A2199 of $\alpha = 16^h28^m47\hbox{$.\!\!^{s}$}0, \delta
= 39^\circ30^\prime22^{\prime\prime}$ (J2000) and $cz = 9156~\kms$.
This position is $3\farcm1 = 80~\kpc$ SE of the X-ray peak.

\begin{figure*}
\figurenum{6}
\label{skyplot}
\epsscale{1.0}
\plotone{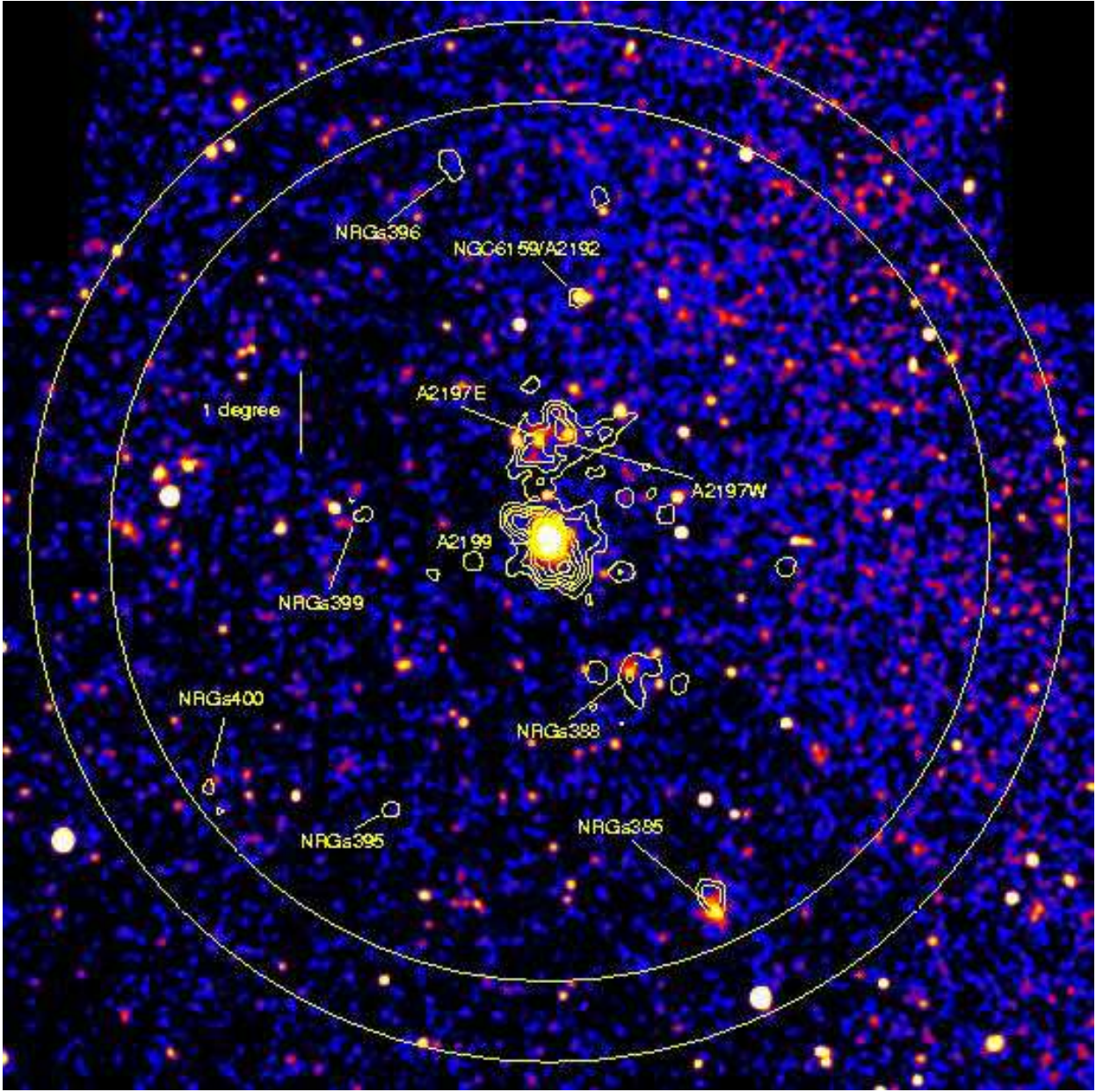} 
\caption{Galaxy density contours overlaid on RASS X-ray data. The RASS
data have been exposure corrected. Both RASS data and the galaxy
density were smoothed with a Gaussian with $2^\prime$ FWHM.  Some
groups of galaxies have no associated X-ray emission detectable in the
RASS. The inner circle is the limit of our redshift survey; our sample
also includes galaxies with redshifts from the literature within the
outer circle.}
\end{figure*}

\begin{figure}
\figurenum{7}
\plotone{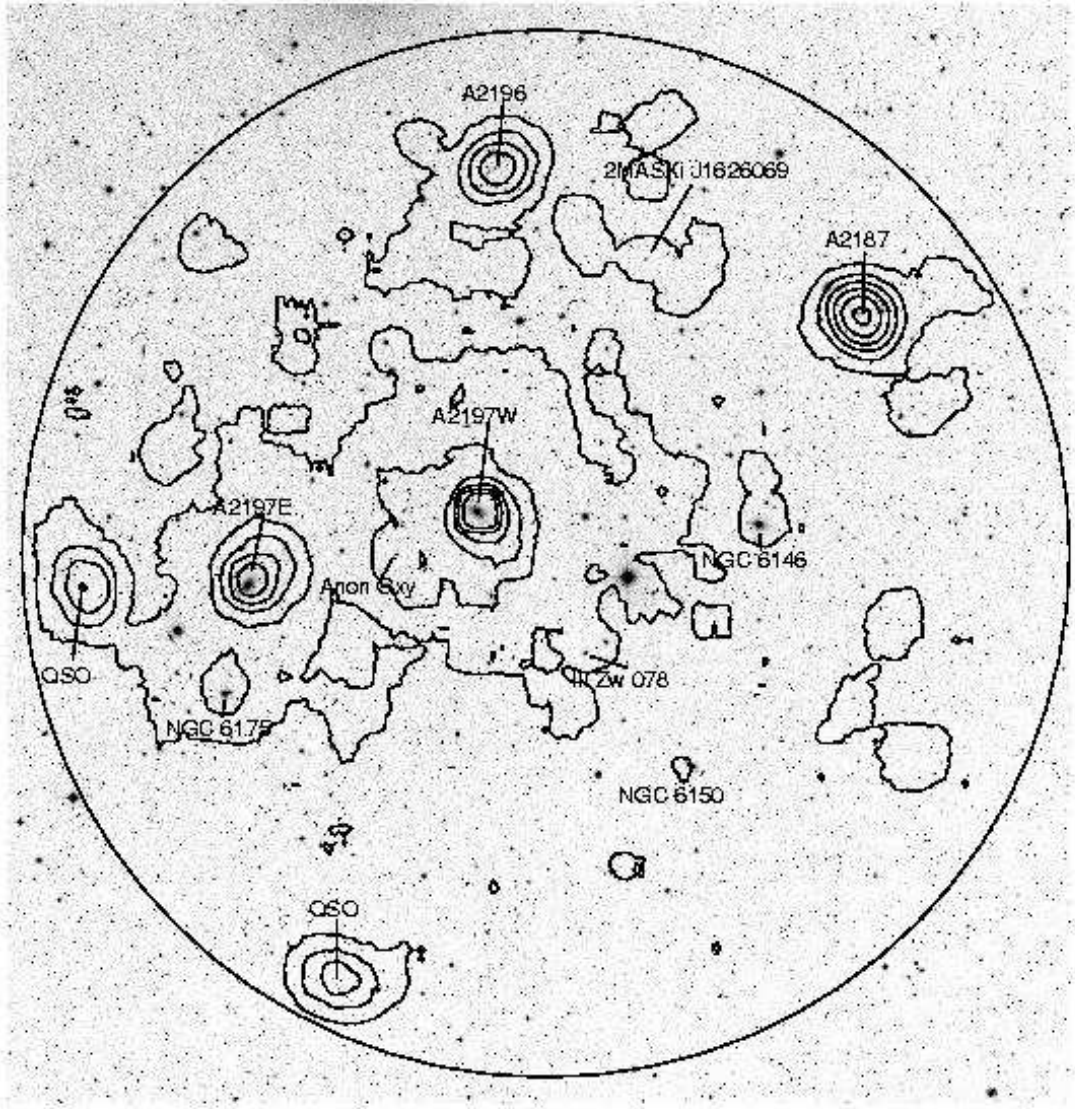} 
\caption{\label{a2197pspc} X-ray contours from a 13.5-ksec {\em ROSAT}
PSPC observation of A2197 overlaid on optical POSS I data.  The
diameter of the PSPC field is $2^\circ$. North is up and East is to
the left.}
\end{figure}

\begin{figure}
\figurenum{8}
\plotone{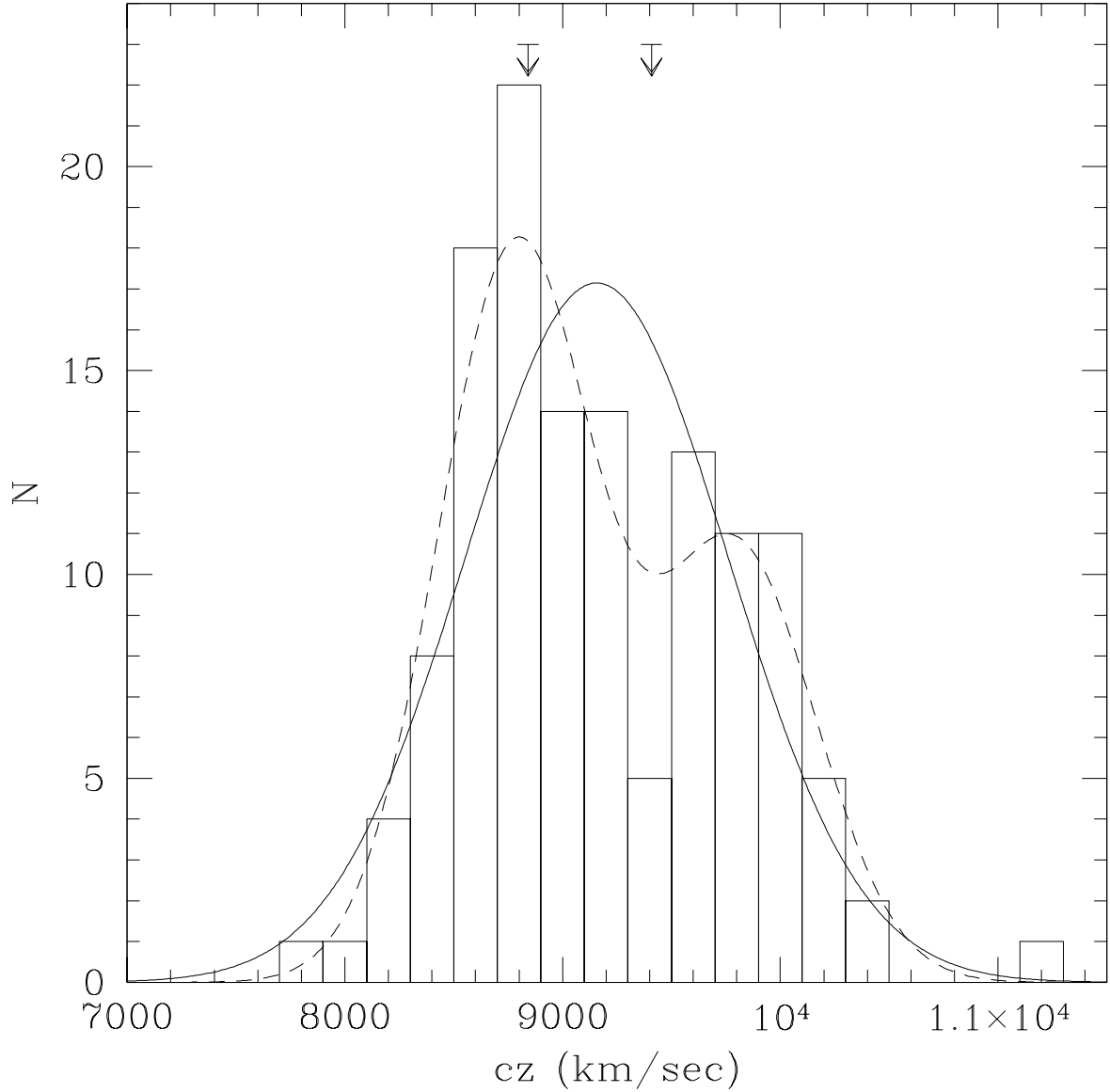} 
\caption{\label{v2197} Redshift histogram of galaxies within
1.5$~\Mpc$ of A2197. The solid line shows the best-fit Gaussian and
  the dashed line shows a two-component model.  Arrows indicate the
  redshifts of NGC 6173 and NGC 6160, the cD galaxies of A2197E and
  A2197W respectively.} 
\end{figure}

\begin{figure}
\figurenum{9}
\epsscale{0.8}
\plotone{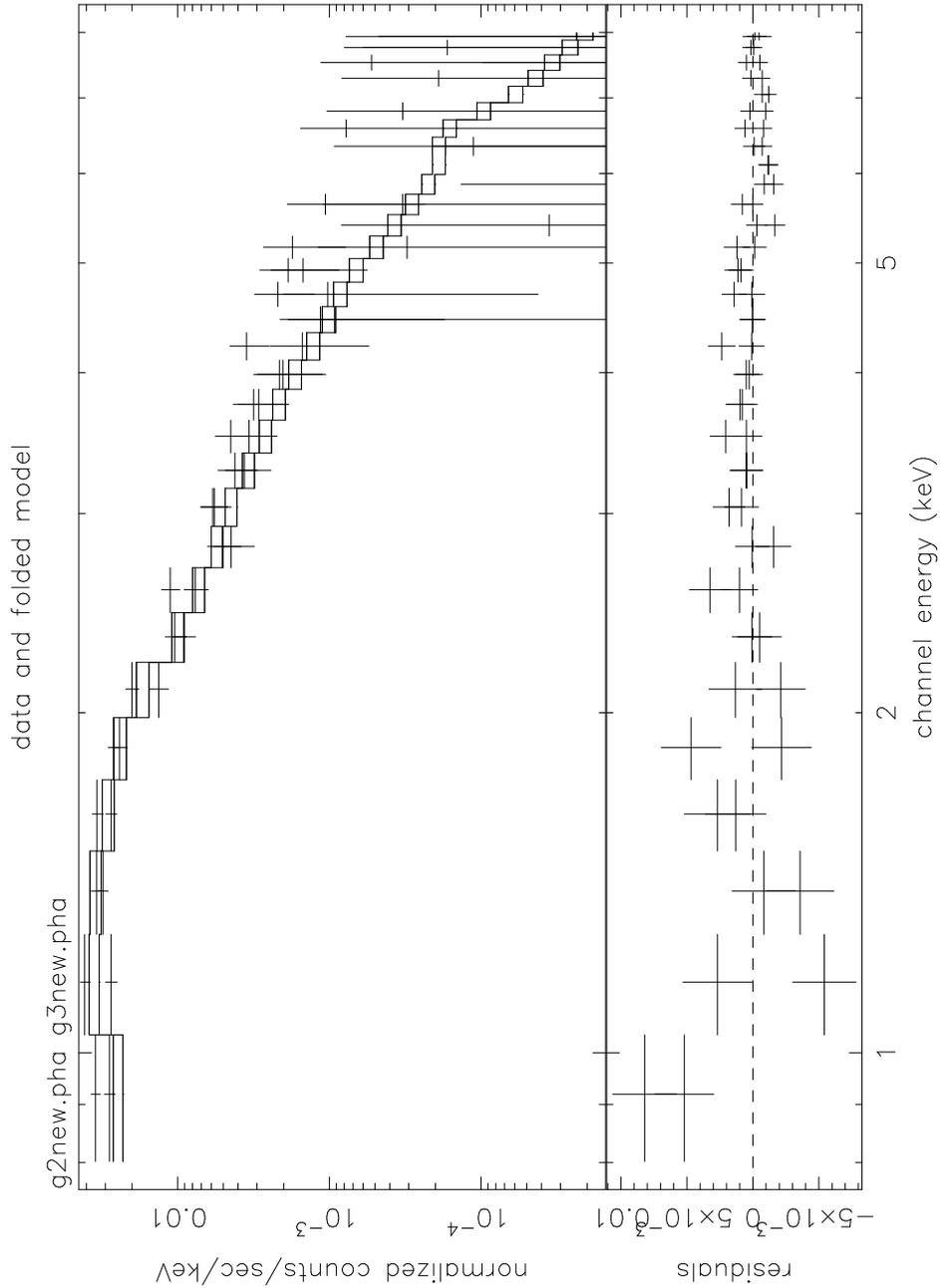} 
\caption{\label{xrayspectrumw} {\em ASCA} X-ray spectrum of
A2197W. The data points show the GIS 2 and GIS 3 instruments and the
lines show the best-fit thermal plasma model convolved with the {\em
ASCA} response. } 
\end{figure}

\citet{rines01b} showed that the sky positions of galaxies in the
infall region reveal several groups in addition to the main cluster.
Figure \ref{skyplot} displays X-ray intensity from the RASS with
contours of the local galaxy density overlaid.  We calculate the
galaxy density contours from cluster members (see definition in $\S
5.7$). Extended X-ray emission usually comes from an intragroup medium
and thus confirms that many, but not all, of the systems are
bound. NRGs395, NRGs399, and NRGs400 are X-ray faint ($L_X<10^{42}
h^{-2} $ergs s$^{-1}$) and thus may not be bound systems; we
drop these from further analysis.  Table \ref{properties} lists the
coordinates and basic properties of the bound systems (see also
MBGR). We calculate redshift centers $cz$ and projected velocity
dispersions $\sigma_p$ according to
\citet{danese} for galaxies within 1.5$~\Mpc$ of the X-ray positions.
Table \ref{properties} lists these properties and the number
$N_{3\sigma_p}$ of member galaxies (within $\pm 3\sigma$ of the
central redshift) from which they are calculated.  Note that the mean
redshift of A2199 is consistent with the hierarchical center, which we
use in all other analyses. A2197W and A2197E overlap both on the sky
and in redshift space (see $\S 4$).  We caution the reader that the
calculated values of $\sigma_p$ for the groups may be unreliable
estimates of their dynamical properties (see $\S 6.1$).

The region around NGC 6159 shows a density enhancement in both X-rays
and galaxy density (visible in Figure 2 of Rines et al.~2001b, but not
discussed there); A2192, a background cluster, lies immediately to the
West. \citet{2000A&A...353..487T} discuss a deep HRI observation of
NGC 6159 (targeted for its unusually high ratio of X-ray to optical
flux); the emission is extended. They suggest that the emission is
intragroup emission but they lack the redshift information necessary
to confirm this suggestion.  Our redshift data show that an optical
group is indeed centered on NGC 6159 and contains as many as 13
members, 7 of which are within 0.5$~\Mpc$ (Table \ref{properties}).

The X-ray emission from A2199 is quite symmetric relative
to other clusters \citep[][MVFS]{mefg}, suggesting that the inner
region of A2199 has not undergone any recent major mergers. The
bound systems are all located at projected radii significantly
larger than the virial radius ($r_{v} \approx 1.6~\Mpc$, see $\S
5.6$). We identify galaxies within $1.5~\Mpc$ of the bound systems in
Figure \ref{caustics}.
NRGs385, NRGs388, and A2199 are roughly colinear, and the galaxy
contours of A2199 are noticeably elongated along this axis (Figure
\ref{skyplot}).  This alignment may be coincidental or it may indicate
the presence of a filament of galaxies and/or dark matter
\citep[see][for further discussion]{rines01b}.  

We next investigate the robustness of the infall mass profile with
respect to variations in the D99 prescription. 

\subsection{Choice of Smoothing Parameter}

The smoothing parameter $q$ defines the width of the smoothing kernel
in angular distance to the smoothing length in redshift.  There is no
simple {\em a priori} definition for this parameter. It is thus
important to quantify the systematic uncertainty due to this choice
\citep[e.g.,][]{rines2000,rines01a}. D99 showed that the choice $q$=25
yields accurate mass profiles for simulated clusters; the choices
$q$=10 and $q$=50 yield similar results. RQCM find that $q$=5 yields a
mass profile for the Shapley Supercluster consistent with the sum of
the X-ray masses of the individual systems.

Figure \ref{mass} displays the mass profile estimated from the
caustics for the choices $q$=10, 25, and 50.  The profiles show
excellent agreement within 4$~\Mpc$, indicating that our results do
not depend strongly on the choice of this parameter. Outside 4$~\Mpc$,
the caustics have negligible amplitude due to the constraint on the
first derivative of the caustics (see D99 and $\S 5.3$).  We adopt the
$q$=25 mass profile as the standard profile to test other
variations. 

\subsection{Defining the Velocity Center of the Supercluster}

In the simplest case, the velocity center of the largest system in a
supercluster (in this case, A2199) is equal to the velocity center of
the entire supercluster.  However, clusters within a particular
supercluster may have substantial peculiar velocities \citep{bahcallsc}.

\citet{lucey97} estimate that the Hubble velocity with respect to the
CMB at the distance of A2199 is $cz_{CMB} = 9190\pm360~\kms$ using the
Fundamental Plane and $9550\pm400~\kms$ using $D_V-\sigma$.  Using the
same data, \citet{hudson97} and \citet{gfb2001} estimate
$9289\pm307~\kms$ and $9452\pm353~\kms$ respectively.  The EFAR survey 
\citep{2001MNRAS.321..277C} reports $cz_{CMB} = 9483\pm741~\kms$ using the
Fundamental Plane.  \citet{giov98}  use the Tully-Fisher relation to
estimate $cz_{CMB} = 9231\pm674~\kms$.  The IR SBF study of
\citet{irsbfh0} suggests $cz_{CMB} = 9302\pm500~\kms$ for NGC 6166,
the cD galaxy of A2199.  The weighted mean yields $cz_{CMB} =
9349\pm157~\kms$ for the Hubble velocity at the distance of A2199,
compared to $cz_{CMB} = cz_\odot + 26~\kms = 9182~\kms$ from our
hierarchical center.  These results suggest that A2199 does not have a
significant peculiar velocity with respect to the CMB.

The systemic redshift $cz_{sup}$ of the supercluster is between 9000
and 9500 $\kms$.  The central cluster A2199 appears to lie at a
slightly lower redshift than the Great Wall.  This offset could mean
that (1) A2199 has a significant peculiar velocity with respect to the
centroid of the supercluster potential or (2) the large-scale
structure of the Great Wall populates only the back side of the infall
region.  In case (1), the redshift center of the supercluster
potential is likely the redshift center of the Great Wall near
A2199. In case (2), however, the supercluster potential is centered on
A2199.

Because the mass profile depends on the squared amplitude of the
caustics, a small shift in the systemic redshift could produce a large
change in the inferred mass profile.  Figures \ref{causcen} and
\ref{vcenmp} show the caustics and resulting mass profiles calculated
using $cz_{sup}=$9000, 9156, and 9500 $\kms$ as the redshift center of
the system and smoothing parameter $q=25$.  The mass profiles for
$cz_{sup}=$9000 and 9156 $\kms$ are in excellent agreement; the mass
inferred for $cz_{sup}=$9500 $\kms$ is somewhat larger outside
$\approx3~\Mpc$, but the uncertainties are correspondingly
larger. Thus, the uncertainty in $cz_{sup}$ introduces $\lesssim 30\%$
uncertainty in the mass within $3~\Mpc$; outside this radius, the
systematic uncertainty introduced is at most a factor of 2.  These
uncertainties are comparable with the intrinsic systematic
uncertainties estimated by D99 for isolated systems.

The X-ray group NRGs385 is located within the supercluster caustics
only for $cz_{sup}=$9500 $\kms$, which is the only variation which
yields a significant caustic amplitude outside $4~\Mpc$.  Indeed, if
one were to sketch the locations of the caustics by eye, one would
likely include the envelope of galaxies outside $4~\Mpc$.  This mass
profile gives a good estimate of the upper bound on the total
supercluster mass.  At the largest radii sampled, the $cz_{sup}=$9500
$\kms$ mass profile yields a total mass of $10^{15}\msun$, roughly
twice the estimates of total supercluster mass within 10$~\Mpc$ for
the standard profile.  The supercluster mass within $8-10~\Mpc$ is
thus more uncertain than the mass within $3-4~\Mpc$.

\subsection{Relaxing the Rules for Determining $\mathcal{A} \mathnormal (R_p)$}

To reduce contamination from interloper galaxies, the method of D99
requires that $d \mbox{ln}\mathcal{A} \mathnormal (r)/ d \mbox{ln} r <
1$. If $d
\mbox{ln}\mathcal{A} \mathnormal (r)/ d \mbox{ln} r > 1$, $\mathcal{A}
\mathnormal (R_p)$ is replaced with a value 
such that $d\mbox{ln}\mathcal{A}\mathnormal(r)/d\mbox{ln}r=1$.
Qualitatively, this step prevents the caustics from ``flaring out''
with radius.  This property is satisfied by all mass profiles with
density profiles decreasing as $\rho \propto r^{-\alpha}$ at large
radii for $\alpha \geq 2$ (all the models considered in $\S 2.1$
satisfy this requirement).  This requirement may introduce a bias into
the final mass profile. Figure \ref{causrelax} shows the shape of the
caustics calculated without this requirement and without requiring
that the caustics be symmetric (i.e., the amplitudes of the upper and
lower caustics are independent).  \citet{rines01b} use this variation.
Figure \ref{massupdown} shows the mass profile calculated from these
caustics using the minimum caustic amplitude at each radius for $q=25$
and $cz_{sup}=9156~\kms$.  Within $4~\Mpc$, the mass profile
calculated in this manner is nearly identical to the standard profile.
Outside $4~\Mpc$, the mass profile lies between the standard profile
and that calculated with $cz_{sup}=$9500~$\kms$.

\subsection{Substructure and Caustics}

In A2199, we directly see the substructure which disrupts the sharp
caustic pattern expected for simple spherical infall \citep{kais87}.
The location of subclusters within the caustic pattern of A2199
demonstrates that they are falling into the supercluster.  We split
the entire region into two halves along a line passing through A2199
to test the importance of A2197 in determining the location of the
caustics.  The caustics at the radius of A2197 have a larger amplitude
towards A2197 than away from it (Figure \ref{ta2197}).  There is a
noticeable deficit of galaxies at the radial distance of A2197 on the
opposite side of A2199.  However, the caustics away from A2197 in the
vicinity of NRGs388 appear similar to those towards A2197, although
the estimator yields a smaller amplitude due to the deficit of
galaxies at smaller radii and the constraint that the caustics not
rise too quickly.  We discuss the interplay of group and supercluster
dynamics in $\S 6$.  Figure \ref{masssplit} shows the mass profiles
calculated towards and away A2197; the two profiles agree within
uncertainties.

\subsection{Comparison to Model Mass Profiles}

We fit the mass profile to the models discussed in $\S 2.1$.  Table
\ref{mpfits} lists the best-fit parameters $a$ (fixed for SIS) and
$M(a)$ for the three models for the variations in $q$, $cz_{sup}$, and
fitting technique described above. The fits only include radii with
$\mathcal{A} \mathnormal (r) > 100~\kms$.  Because the data points in
the mass profile are not independent, the absolute values of $\chi_\nu
^2$ listed in Table \ref{mpfits} are only indicative, but it is clear
that, as for the simulated superclusters, the NFW and Hernquist
profiles provide acceptable fits to the infall mass profile while the
SIS is excluded for all variations. The NFW profile generally provides
a better fit to the data than the Hernquist profile. The parameters of
the best-fit mass profile are quite robust with respect to the
variations in the estimator discussed above.  As noted in $\S 4.2$,
the mass profile calculated assuming $cz_{sup}=$9500 $\kms$ yields a
significantly larger total mass than the other estimates, but the mass
profile inside $3~\Mpc$ is very similar to the others.

For the standard mass profile, the best-fit NFW parameters are $\mbox{log}
~\delta_c = 4.4$, concentration $c = r_{200}/a = 8$, and
$M_{200}=3.2\times 10^{14} = 7.8 M_*$ for the NFW definition of $M_*$
in a $\Lambda$CDM model ($M_{200}=20 M_*$ for the SCDM model in
NFW). Our results match the predicted $\delta_c-M_{200}$ relation for most of the
cosmological models in NFW.

\subsection{Comparison to Other Mass Estimates}

MVFS use {\em ROSAT} X-ray images and spatially-resolved spectroscopy
from {\em ASCA} to estimate the mass profile of A2199 in the range
$0.1-0.6~\Mpc$. This mass profile is one of the few X-ray cluster mass
profiles measured to such a large radius with a resolved temperature
profile.  The X-ray mass profile, which is completely independent of
the caustic technique, shows excellent agreement with the standard
profile (Figure \ref{massx}).  Figure \ref{massx} also shows a mass
estimate from deprojecting {\em Einstein} data \citep{wjf97} and the
best-fit NFW profile to the deprojected {\em Chandra} data at small
scales \citep[][however, note that this NFW profile is formally
rejected by the data]{jafs}.  The standard mass profile agrees well
with the former but not with the latter.  The ICM is far from
hydrostatic equilibrium in the core of A2199, so the mass estimate at
these small radii is perhaps not reliable.

We use the prescription of \citet{girardi98} to estimate a virial
radius $r_{v}=1.6~\Mpc$ for A2199.  We estimate the virial mass of
A2199 using all galaxies inside the caustics and within $r_v$.  The
virial mass is $5.34\times 10^{14}\msun$, in good agreement with
\citet{girardi98}.  The standard profile yields a mass about 30\%
smaller than this value at $r_v$. This difference is within the
uncertainty range expected for the caustic technique.

\subsection{Velocity Dispersion Profile}

As a final consistency check of the caustic mass profile, we calculate
the velocity dispersion as a function of radius in bins of 25
galaxies. For this calculation, we include all galaxies inside the
caustics of either the standard caustics or the caustics calculated
with $cz_{sup}=9500~\kms$.  Including only galaxies in the former
caustics yields similar results within $4~\Mpc$; very few galaxies are
classified as members beyond this radius.  Figure \ref{sigmar}
displays both the regular and the integrated velocity dispersion
profiles along with the predicted profiles based on the best-fit
Hernquist and NFW mass profiles and an assumption of isotropic orbits.
The predicted profiles match the observed profiles quite well, and
thus the galaxy orbits in A2199 are consistent with isotropic.  Note
that, as in A576, the integrated velocity dispersion decreases
smoothly with radius and does not reach an asymptotic value which
could be easily interpreted as the velocity dispersion of the cluster.

\section{Discussion}

Projection effects introduce the largest uncertainty in the caustic
technique both for isolated clusters and simple superclusters (D99).
The X-ray masses for A2197W and A2197E are factors of 6 and 13 smaller
than A2199.  Tests of similar systems from simulations indicate that
the caustic technique should yield an accurate mass profile for the
supercluster.  The uncertainty in the redshift center of the
supercluster results in a factor of two uncertainty in the total
supercluster mass within 8-10$~\Mpc$. Within $3~\Mpc$, however, we
estimate the error on the enclosed mass to be only about 30\%,
comparable to the intrinsic uncertainty estimated by simulations.

\subsection{Do Supercluster Dynamics Dominate Group Dynamics?}

At the projected radii of A2197W and A2197E, the inferred group masses
are a small fraction of the enclosed supercluster mass.  The
supercluster dynamics may dominate the observed dynamics of groups in
the infall region.  The supercluster may increase the velocity
dispersion of the groups either directly or by introducing interlopers
which are bound to the supercluster but are not group members.  

One test of the relative importance of group and supercluster dynamics
is to predict the velocity dispersion of the groups from their X-ray
luminosities or temperatures and the $L_X - \sigma$ or $\sigma - T_X$
relation \citep[][MBGR and references therein]{2000ApJ...538...65X}.
We list the predicted and observed velocity dispersions in Table
\ref{lxsig}.  Except for A2199 and the NGC 6159 group, the observed
velocity dispersions are larger than those predicted by the
$L_X-\sigma$ or the $\sigma - T_X$ relation, suggesting that the
observed dynamics of the groups are dominated by the supercluster,
either physically (e.g., due to tidal forces from filaments) or
observationally due to the increased density of interlopers.

At the radii of A2197, NRGs388, and NGC 6159, the amplitude of the
caustics is relatively unaffected by the presence of these subsystems.
At the projected distance of NRGs385, however, the caustic pattern
(Figure \ref{causrelax}) and the velocity dispersion profile (Figure
\ref{sigmar}) shows a larger amplitude than at smaller radii. This
result shows that the caustic amplitude is strongly affected by
substructure at this radius. The redshift center of the supercluster
determines whether NRGs385 is inside the supercluster caustics.

\subsection{X-ray Contribution of Radio Sources}

X-ray emission from radio sources can contribute significantly to the
total X-ray luminosity of a galaxy group. Indeed, \citet{feretti} find
that both NRGs385 and NRGs388 contain radio galaxies which are
detected in X-rays.  From {\em ROSAT} PSPC data, they find that NGC
6107 and NGC 6109 contain 6\% and 1.8\% respectively of the total
X-ray luminosity of NRGs385.  For NRGs388, they suggest that the
observed X-ray luminosity is entirely associated with the halo of NGC
6137 and not with an intragroup medium.  Reliable estimates of
$\sigma_p$ from the $L_X - \sigma_p$ relation therefore require
pointed X-ray observations of groups to determine the X-ray luminosity
due to the intragroup medium.

\subsection{Are the Groups Really Infalling?}

The location of the groups within the caustics indicates that they are
dynamically linked to the A2199 supercluster.  This connection,
however, does not necessarily imply that the groups are currently
infalling onto the supercluster.  The turnaround radius $r_{ta}$ for a
$\Lambda =0$ universe satisfies $\rho_m (<r_{ta}) = \Omega_m \rho_c (1 +
\Delta_{0,turn})$ where $\rho_c$ is the critical density and
$\Delta_{0,turn}$ is given by Equation 8 in \citet{rg89}.  In the
limit $\Omega_m=1$, $\rho_m (<r_{ta}) = (9 \pi^2/16)\rho_c$.  The
standard mass profile yields $r_{ta}\approx$4.5-5.7$~\Mpc$ for
$\Omega_m = 1 \rightarrow 0.025$ where we take a lower limit on
$\Omega_m$ from the baryon density $\Omega_b h^2 = 0.025$
\citep{2001ApJ...560...41P} and the assumption that $h \leq 1$.  For
the more massive profile inferred by assuming $cz_{sup}=$9500$~\kms$,
the turnaround radius for the above range of $\Omega_m$ is
$r_{ta}=$5.0-6.5$~\Mpc$. The turnaround radius is smaller in the
presence of a non-zero cosmological constant, but this effect is small
in the present epoch for $\Lambda \lesssim 1$
\citep{1991ApJ...374...29L,1991ApJ...377....7M}.  
Thus, A2197W, A2197E, and NRGs388 are in the infall region of A2199
for most combinations of geometry and cosmology (the expectation value
of the deprojected radius is $<r> = \pi R_p/2$).  The NGC 6159 group
is a borderline case; it is infalling for the standard mass profile
and our adopted cosmology only for small angles ($\lesssim 20^\circ$)
between the plane of the sky and the line connecting A2199 and NGC
6159.  NRGs385 and NRGs396 are not infalling onto the supercluster.  A
simulation of the future evolution of the nearby universe \citep{nl02}
shows that some objects outside the turnaround radius of clusters in
the present epoch but with large peculiar velocities may eventually
become cluster members.

Based on the presence of a caustic pattern around NRGs385 and an
unusual offset between its X-ray and optical peaks, we previously
suggested that NRGs385 might be infalling onto A2199 \citep{rines01b}.
We have shown here, however, that the mass profile robustly excludes
this possibility.  The unusual morphology of NRGs385 is probably
related to the local large-scale structure and is not caused by
effects related to infall (e.g., ram pressure stripping).

The caustic pattern is present beyond the turnaround radius,
suggesting that the gravitational potential of a cluster (or simple
supercluster) and surrounding large-scale structure can significantly
affect the kinematics of objects outside the infall region.  The
caustics extend beyond the turnaround radius in the simulations of
D99. A similar effect is seen in the galaxy-galaxy autocorrelation
function \citep[e.g.,][]{2001Natur.410..169P}.  The existence of
large-scale structure obscures the location of the turnaround radius
in the projected radius-redshift diagram.

\section{Conclusions}

We have previously demonstrated that the infall region of A2199
contains several bound subclusters \citep{rines01b}.  Here, we show
that, in simulations, caustics yield accurate mass profiles of 
superclusters dominated by a central cluster 2-6 times more massive
than all other systems in the supercluster.  The mass profiles of the
simulated clusters agree with NFW or Hernquist models but not with a
singular isothermal sphere.

X-ray data demonstrate that A2197 is the superposition of two bound
X-ray groups (centered on the bright galaxies NGC 6160 and NGC 6173)
which may be interacting.  Because of the complexity of A2197, the
virial theorem does not yield an accurate mass estimate.  The X-ray
masses for A2197W and A2197E are factors of 6 and 13 smaller than
A2199.  Tests of similar superclusters from simulations indicate that
the caustic technique should yield an accurate mass profile for the
A2199 supercluster.

The supercluster mass profile suffers some uncertainty due to the
complexity of the surrounding large-scale structure and the resulting
difficulty in determining the redshift center of the supercluster.
Within $3~\Mpc$, however, this uncertainty is no larger than the
scatter predicted by numerical simulations for isolated clusters. The
infall mass profile is in excellent agreement with independent X-ray
estimates at small radii. At large radii, the mass profile is clearly
inconsistent with an isothermal sphere but agrees with profiles
motivated by simulations.

The mass profile indicates that A2197W, A2197E, and NRGs388 are bound,
infalling groups; NRGs385 and NRGs396 (and the NGC 6159 group for
inclination angles $\gtrsim 20^\circ$) are not infalling onto the
supercluster in the present epoch.  This result demonstrates that the
presence of caustics at large radii from clusters does not prove that
galaxies or groups within these caustics are inside the infall region.
In other words, the gravitational potential of a cluster and
surrounding large-scale structure can significantly affect the
kinematics of objects outside the turnaround radius.  The observed
velocity dispersions of the groups surrounding A2199 are significantly
larger than predicted by their X-ray properties and scaling
relations.  The observed kinematic properties of galaxy groups within
and near superclusters are likely dominated by the supercluster and
not by the group potential.

The caustic technique is a robust estimator of cluster mass profiles
ranging in mass from Fornax to the Shapley Supercluster.  We
demonstrate that the technique is robust for simple superclusters,
both in simulations and in observations of the A2199/A2197
supercluster.  We plan to analyze a sample of 8 nearby clusters and
study the mass-to-light ratio throughout the infall regions of several
of these systems.

\acknowledgements

This project would not have been possible without the assistance of
Perry Berlind and Michael Calkins, the remote observers at FLWO, and
Susan Tokarz, who processed the spectroscopic data.  KR and MJG are
supported in part by the Smithsonian Institution. JJM is supported by
NASA LTSA grant NAG 5-11415. The N-body simulations, halo and galaxy
catalogues used in this paper are publically available at
http:://www.mpa-garching.mpg.de/NumCos.  The simulations were carried
out at the Computer Center of the Max-Planck Society in Garching and
at the EPCC in Edinburgh, as part of the Virgo Consortium project. We
also thank the Max-Planck Institut f\"ur Astrophysik where some of the
computing for this work was done.  The National Geographic Society -
Palomar Observatory Sky Atlas (POSS-I) was made by the California
Institute of Technology with grants from the National Geographic
Society.  This research has made use of the NASA/IPAC Extragalactic
Database (NED) which is operated by the Jet Propulsion Laboratory,
California Institute of Technology, under contract with the National
Aeronautics and Space Administration. We thank the referee for
detailed comments which improved the presentation of this paper.

\bibliography{rines}

\clearpage

\clearpage

\begin{figure}
\figurenum{10}
\epsscale{0.8}
\plotone{Rines.fig10.ps} 
\caption{\label{xrayspectrume} Same as Figure \ref{xrayspectrumw} but
for A2197E. } 
\end{figure}

\begin{figure}
\figurenum{11}
\epsscale{1.0}
\plotone{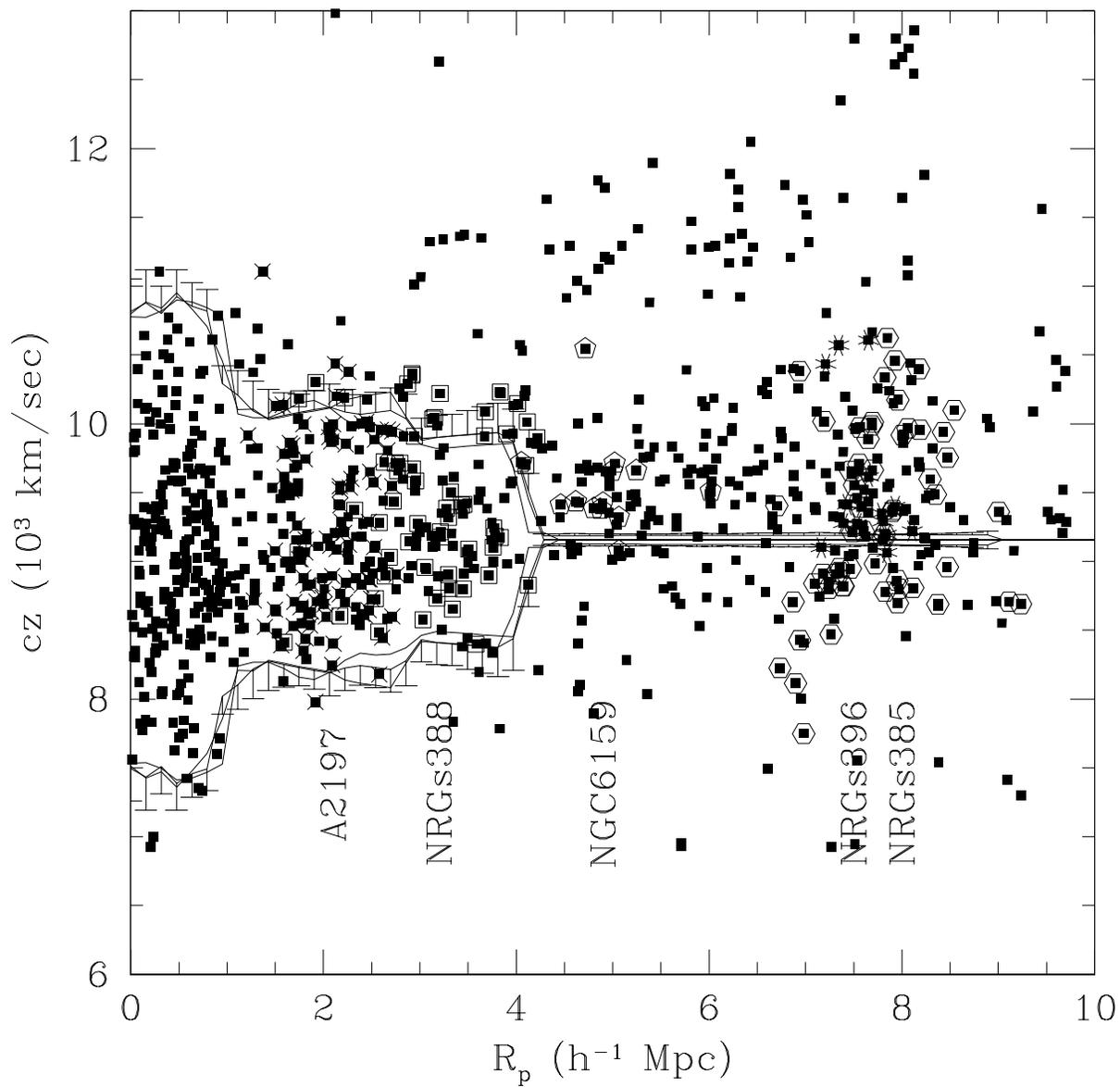} 
\caption{\label{caustics} Projected radius versus redshift for
galaxies surrounding A2199.  Lines indicate our estimate of the
caustics.  Crosses, hexagons, open squares, asterisks, and pentagons
indicate galaxies in A2197, NRGs385, NRGs388, NRGs396, and NGC6159
respectively.}
\end{figure}

\begin{figure}
\figurenum{12}
\plotone{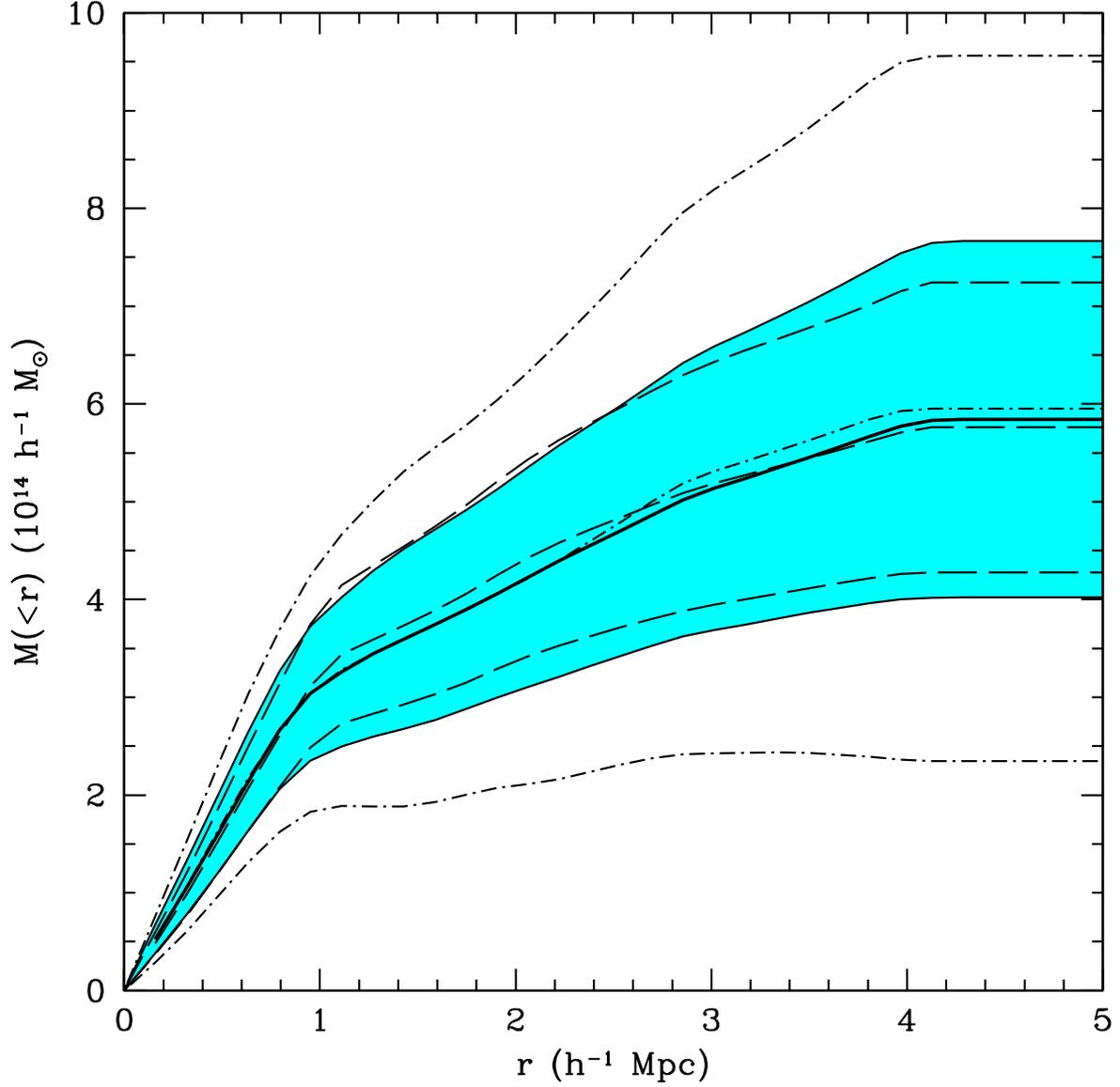} 
\caption{\label{mass} Mass profile of the A2199 supercluster. The
thick solid line indicates the mass profile estimated using $q=25$
(see text for definition), the thin solid lines  and shaded region
show the 1-$\sigma$ 
uncertainties in this profile.  The sets of dashed and dash-dotted
lines indicate the mass profiles and 1-$\sigma$ uncertainties for
$q$=10 and 50 respectively.} 
\end{figure}

\begin{figure}
\figurenum{13}
\plotone{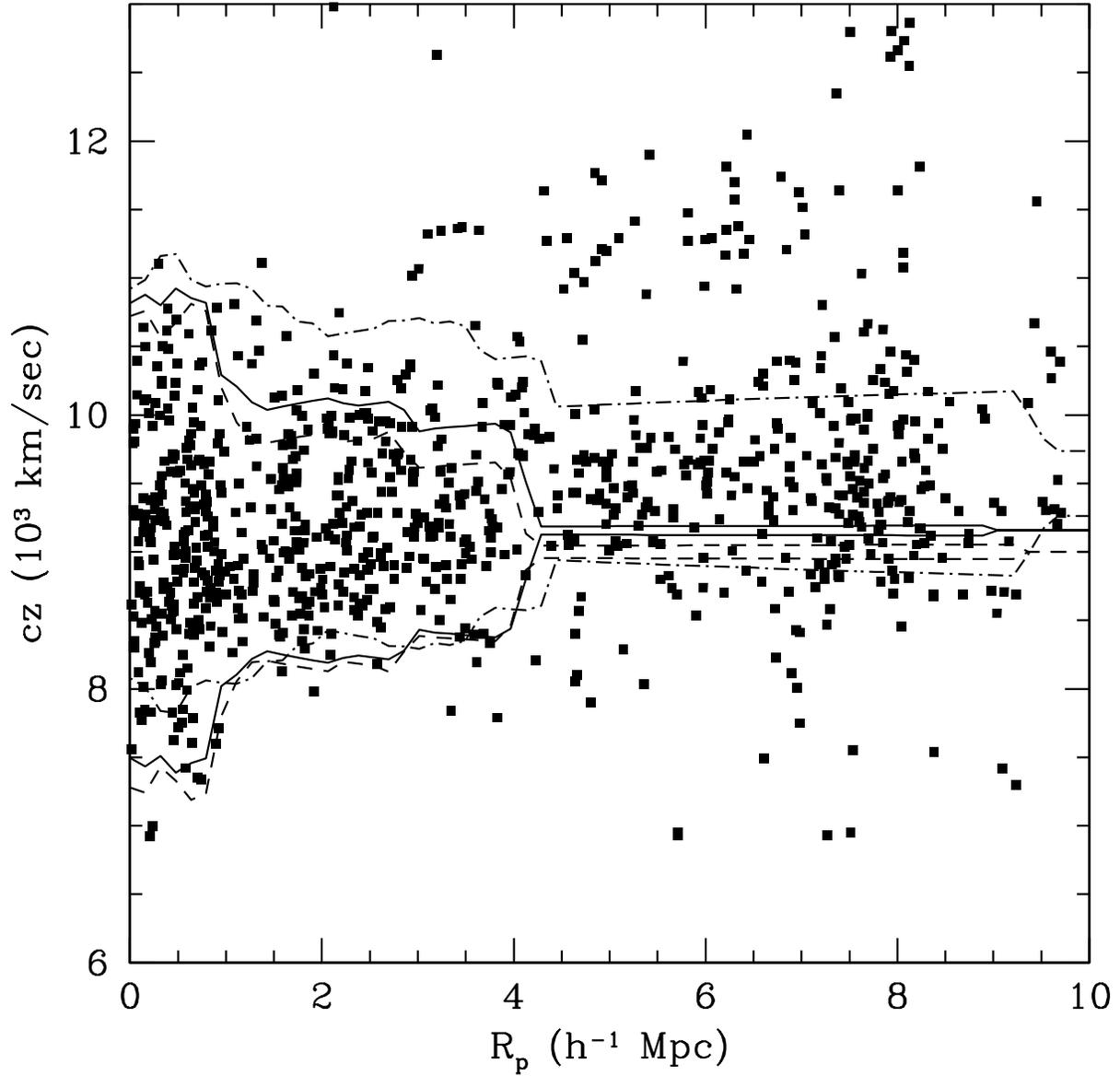} 
\caption{\label{causcen} Same as Figure \ref{caustics}, but varying
the systemic redshift of the supercluster $cz_{sup}$.   Lines indicate
our estimate of the caustics.  Dashed, solid, and dash-dotted lines
show the caustics calculated with $cz_{sup}$ = 9000, 9156, and
9500$~\kms$ respectively.}
\end{figure}

\begin{figure}
\figurenum{14}
\plotone{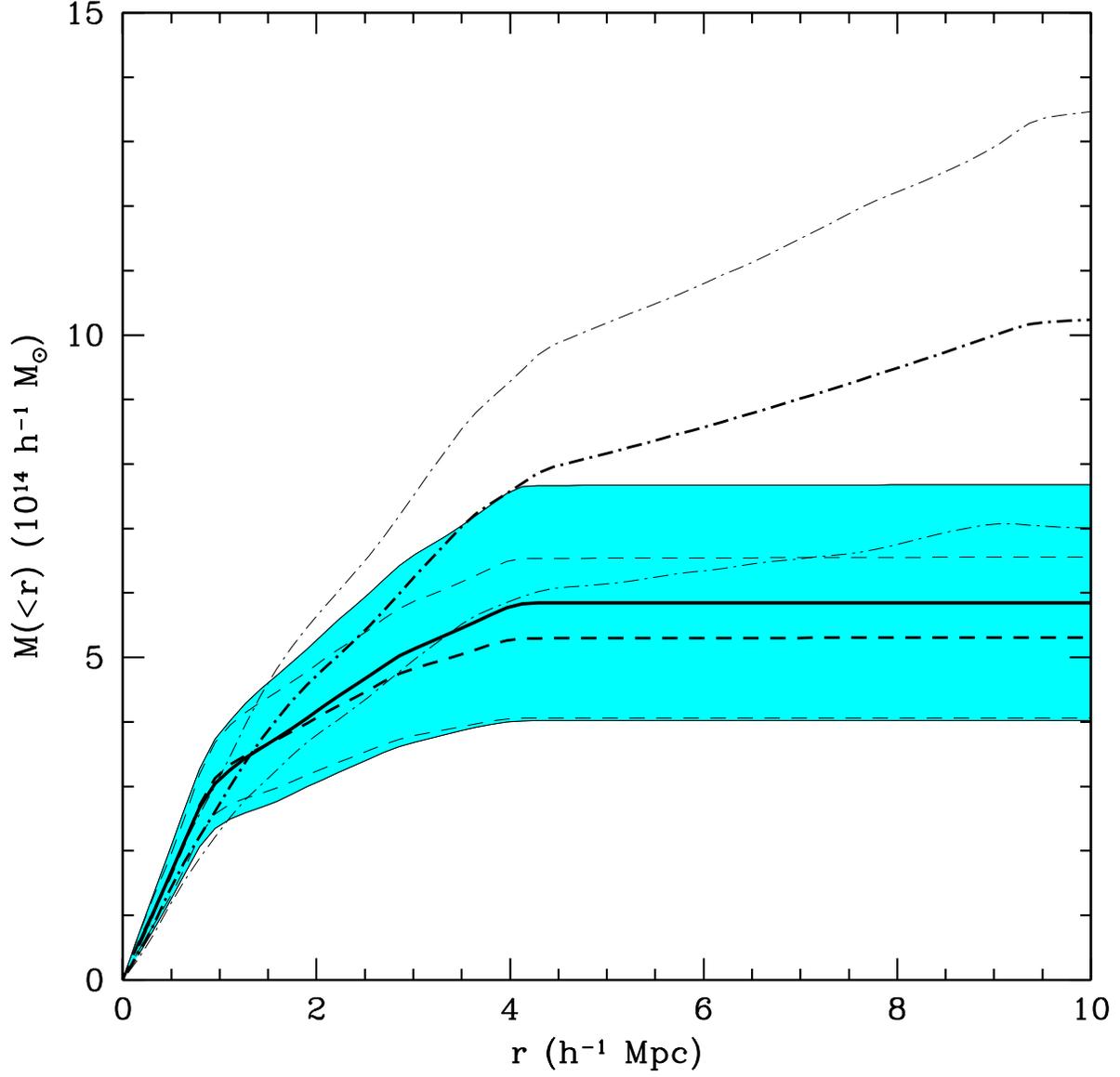} 
\caption{\label{vcenmp} Effect on the mass profile of varying the
central redshift of the supercluster.  Dashed, solid, and dash-dotted lines
show the caustics calculated with $cz_{sup}$ = 9000, 9156, and
9500$~\kms$ respectively.} 
\end{figure}

\begin{figure}
\figurenum{15}
\plotone{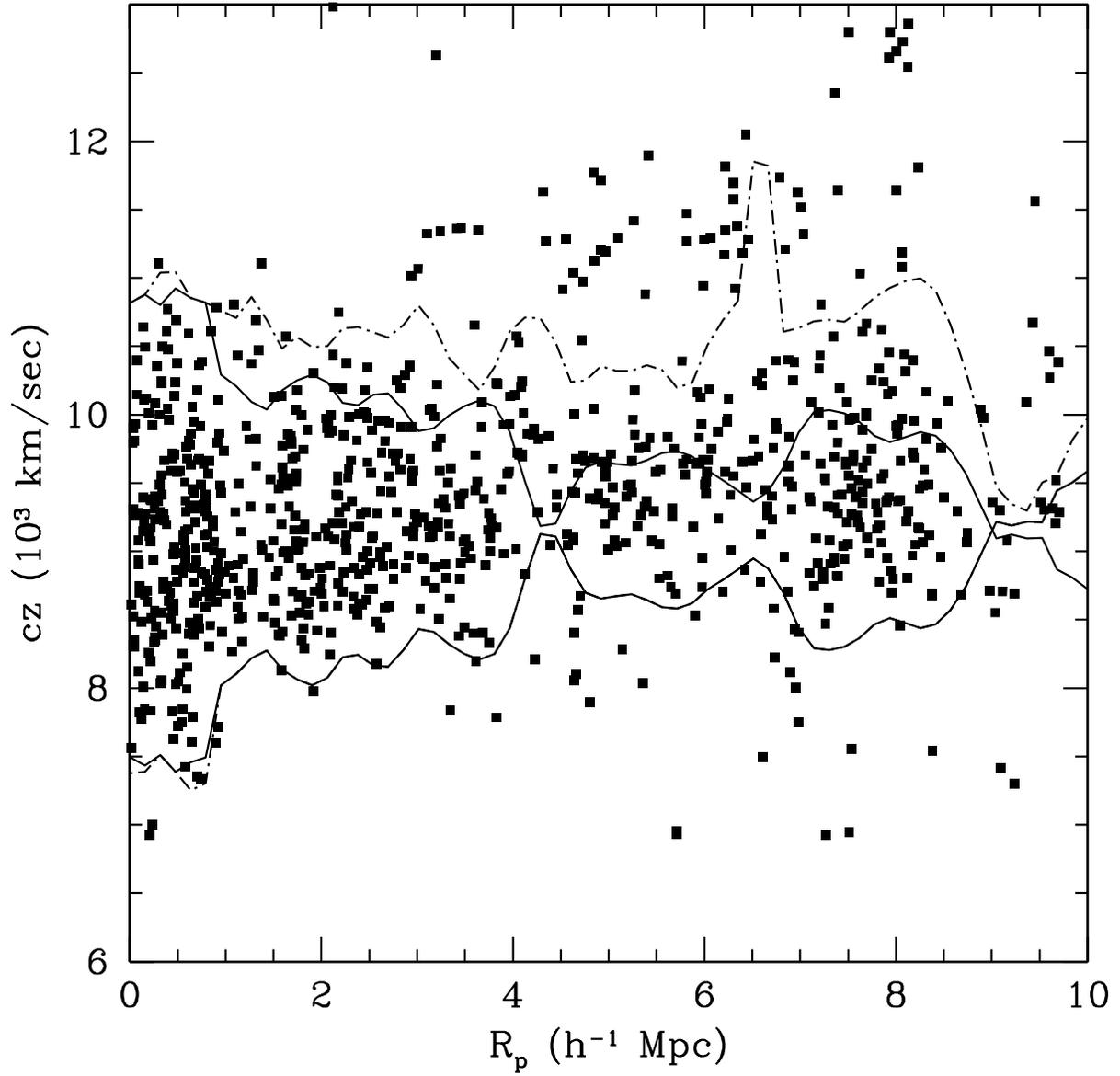} 
\caption{\label{causrelax} Same as Figure \ref{caustics}, but using
relaxed requirements to calculate the caustics.  Dashed lines show the
caustics on either side of the cluster; the solid lines show the
minimum caustic amplitude at each radius. The latter caustics are
defined primarily by the lower caustic.}
\end{figure}

\begin{figure}
\figurenum{16}
\plotone{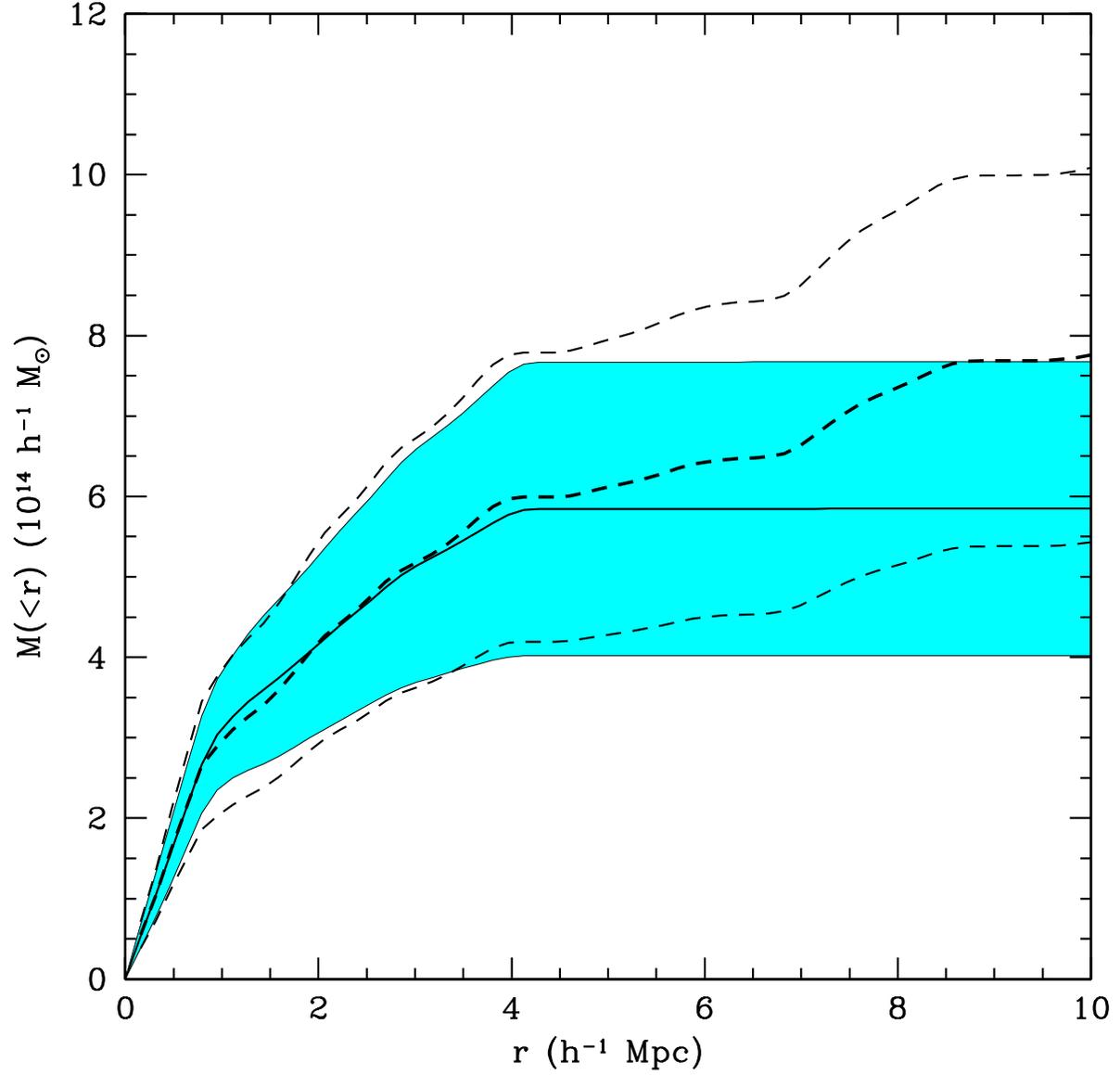} 
\caption{\label{massupdown} Mass profile calculated with relaxed
requirements on the slope of the caustics. The dashed lines show the
standard caustics for reference.} 
\end{figure}

\begin{figure}
\figurenum{17}
\plotone{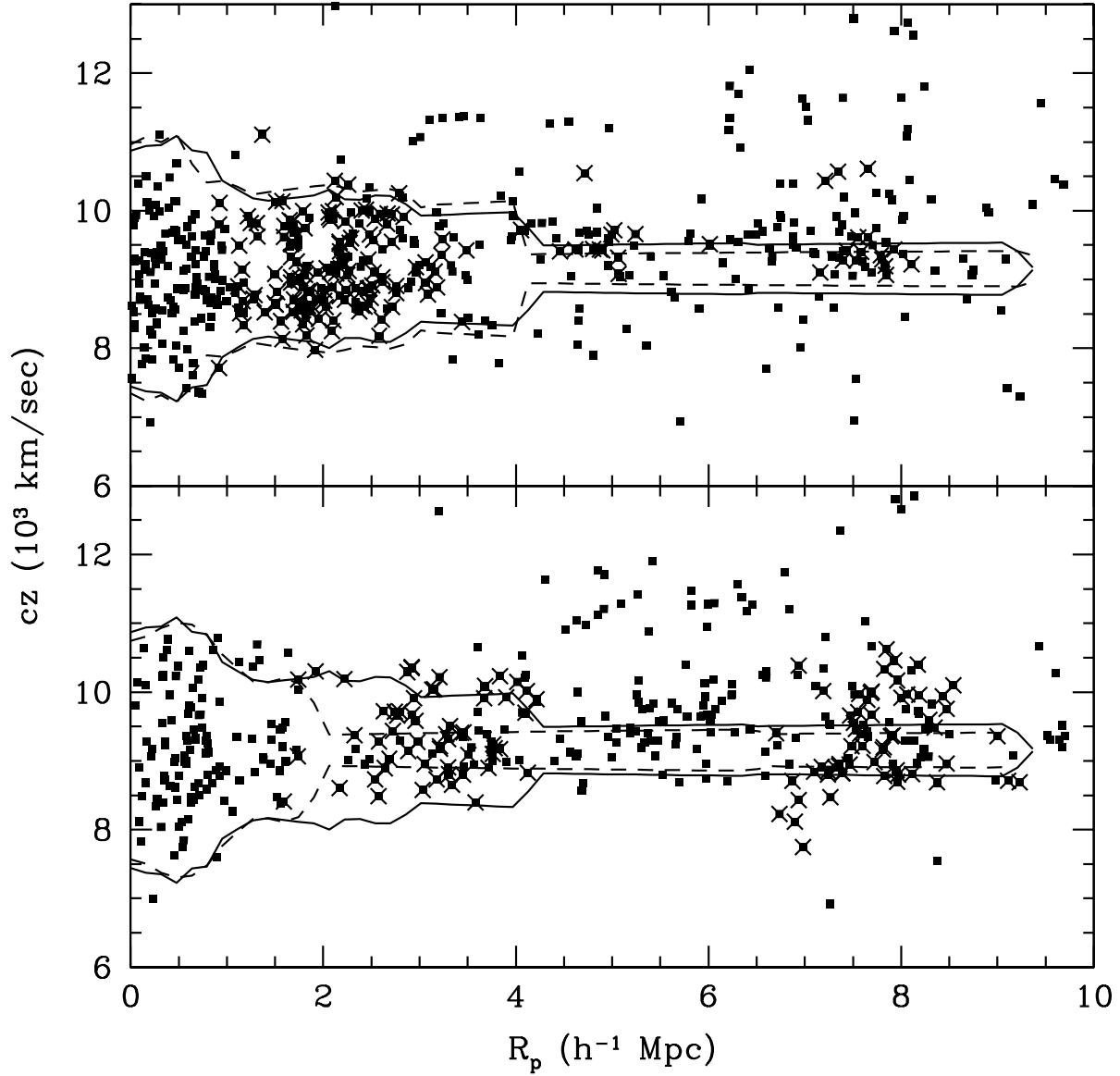} 
\caption{\label{ta2197} Projected radius versus redshift for
galaxies surrounding A2199.  The top panel shows galaxies towards
A2197; the bottom panel shows galaxies away from A2197.  Dashed lines
indicate the caustics calculated for each sample; the solid lines are
the caustics from the full sample.  Crosses indicate galaxies in
A2197, NGC 6159, and NRGs396 (top panel) and NRGs385 and NRGs388
(bottom panel).}
\end{figure}

\begin{figure}
\figurenum{18}
\plotone{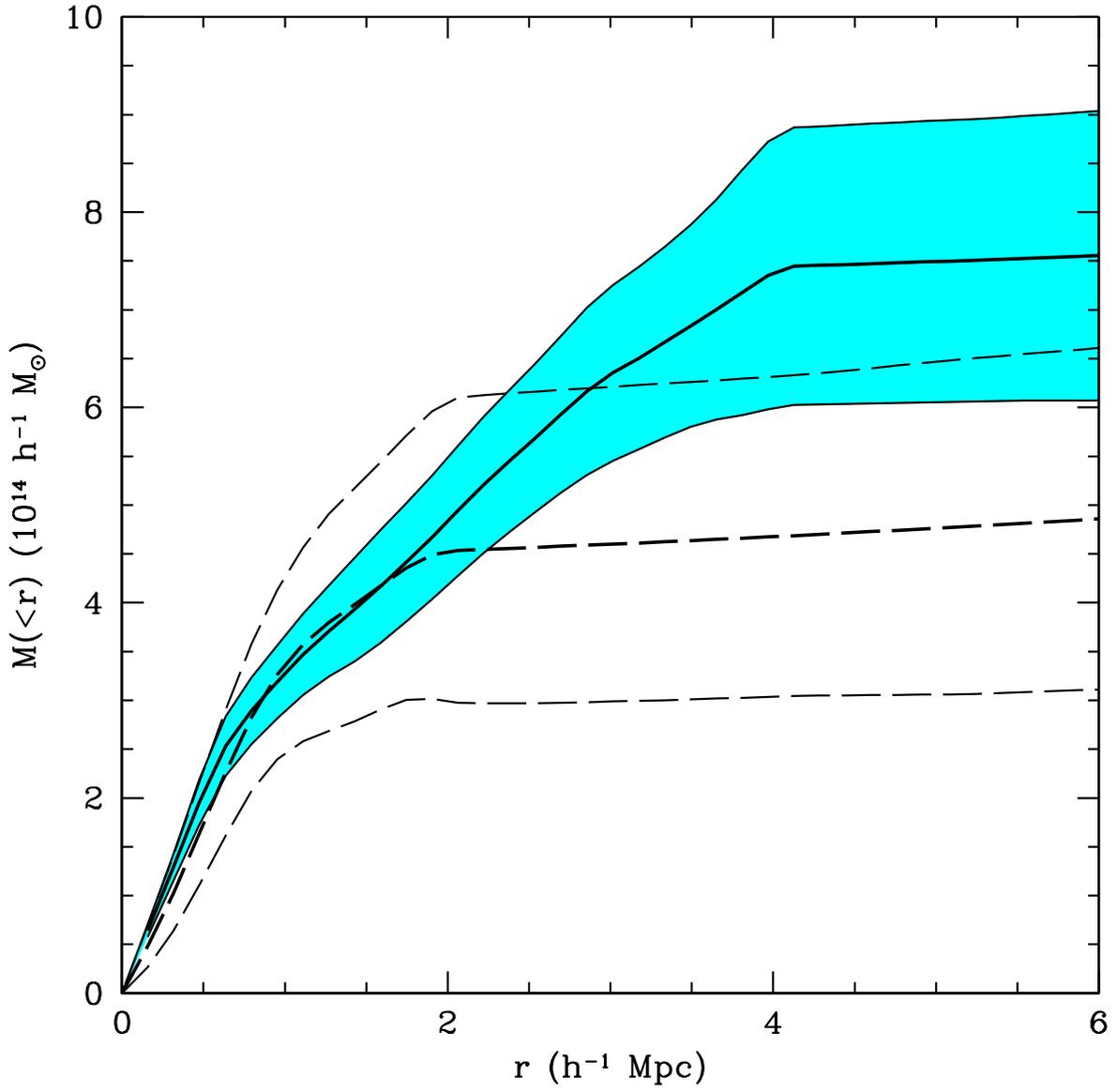} 
\caption{\label{masssplit} Mass profile calculated towards (solid
lines) and away (dashed lines) from A2197.} 
\end{figure}

\begin{figure}
\figurenum{19}
\plotone{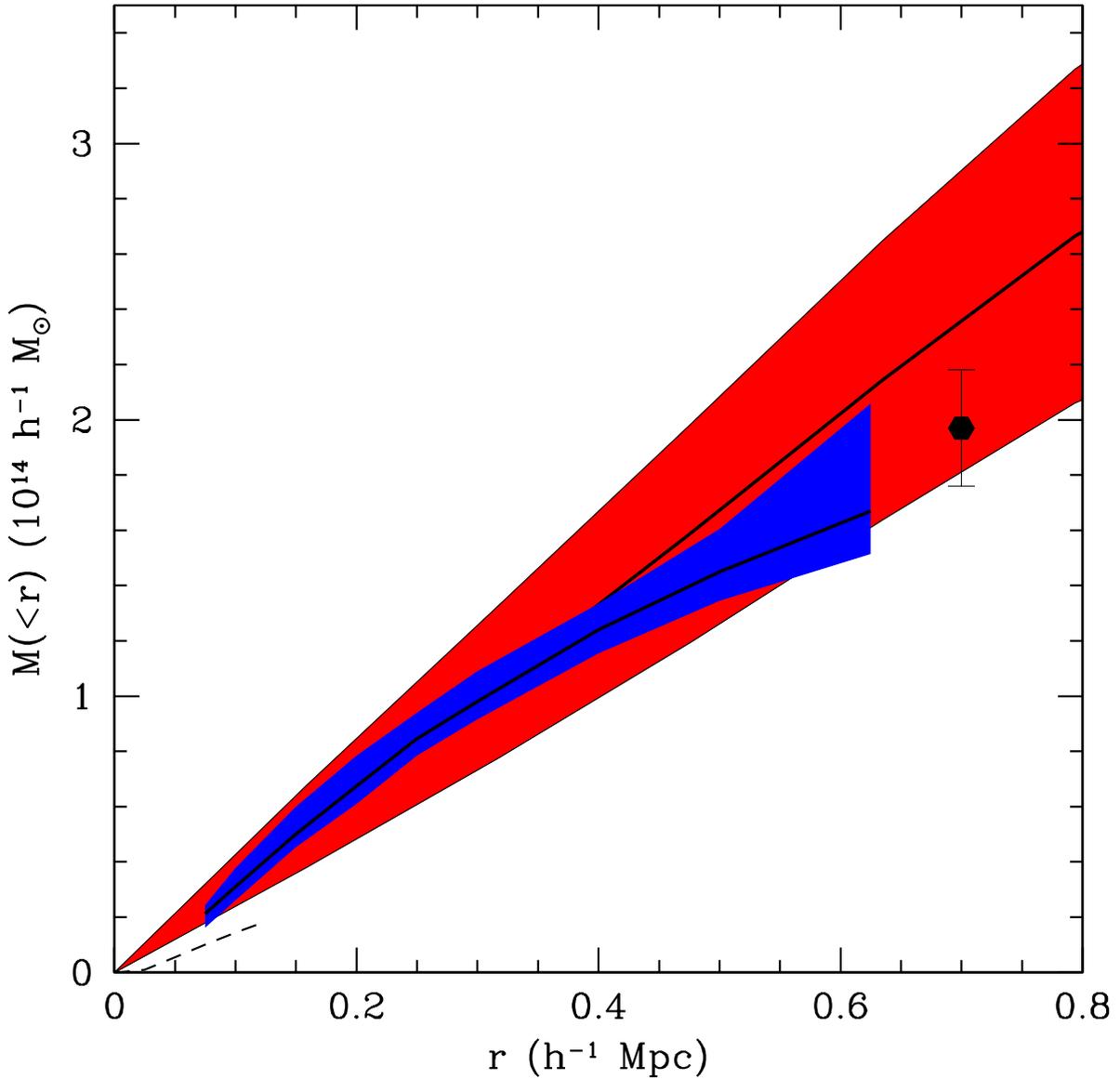} 
\caption{\label{massx} Comparison of infall mass profile to
independent X-ray estimates. The 1-$\sigma$ range of the infall mass
profile is shown in red and the 90\% confidence range of the X-ray
mass profile is shown in blue.  The NFW profile which best fits the
X-ray data is indicated by a thick solid line. The point indicates a
deprojected X-ray mass estimate and the dashed line shows the NFW
profile which best fits {\em Chandra} X-ray data.} 
\end{figure}

\begin{figure}
\figurenum{20}
\plotone{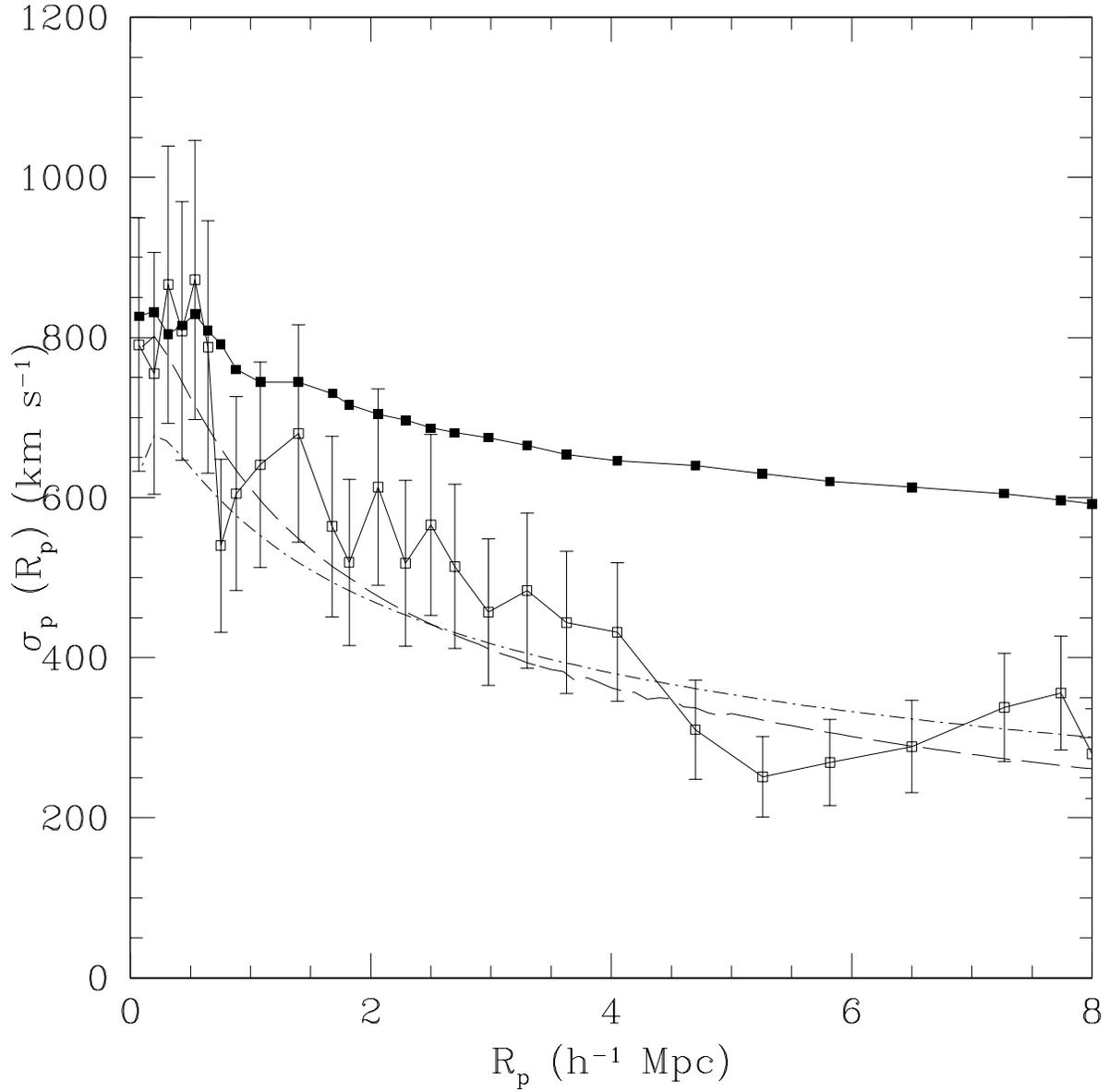} 
\caption{\label{sigmar} Velocity dispersion profile of A2199 member
galaxies (open squares). Filled squares show the integrated velocity
dispersion profile. Dotted and dot-dashed lines show the profiles
predicted by the Hernquist and NFW profiles respectively assuming
isotropic orbits. The predicted profiles have no free parameters.} 
\end{figure}

\clearpage

\begin{table*}[th] \footnotesize
\begin{center}
\caption{\label{sims} \sc Properties of Simulated Simple Superclusters}
\begin{tabular}{lcccc}
\tableline
\tableline
\tablewidth{0pt}
System & $M_1$ & $M_1/M_2$ & $M_{subhalo}$ & $R_{subhalo}$  \\ 
 & $10^{13}\msun$ &  &  $10^{13}\msun$ & $\Mpc$  \\
\tableline
121 & 18.2 & 1.95 & 9.3 & 3.19 \\
 & & & 8.8 & 3.55 \\
 & & & 4.1 & 4.96 \\
 & & & 4.7 & 5.87 \\
156 & 33.5 & 4.43 & 7.6 & 2.90 \\
 & & & 2.1 & 3.66 \\
 & & & 1.2 & 6.74 \\
 & & & 2.5 & 10.13 \\
162 & 31.0 & 6.20 & 3.2 & 2.74 \\
 & & & 5.0 & 3.38 \\
 & & & 1.1 & 7.97 \\
 & & & 2.0 & 8.89 \\
\tableline
\end{tabular}
\end{center}
\end{table*}

\begin{table*}[th] \footnotesize
\begin{center}
\caption{\label{simmpfits} \sc Simulated Mass Profile Fit Parameters}
\begin{tabular}{llcccc}
\tableline
\tableline
\tablewidth{0pt}

System & Profile   &$a$  & $M(a)$  & $\chi ^2$ & $\nu$ \\
& &  $\Mpc$  & $10^{14} M_\odot$ & & \\
\tableline 
121 & NFW & 0.24$\pm0.01$  & 0.38$\pm0.02$  & 101 & 23 \\
121 & Hernquist & 0.49$\pm0.02$  & 0.98$\pm0.05$  & 167 & 23  \\
121 & SIS & 0.5  & 0.42$\pm0.01$  & 1152 & 24 \\
\tableline
156 & NFW  & 0.18$\pm0.01$  & 0.48$\pm0.02$  & 18 & 23 \\
156 & Hernquist  & 0.41$\pm0.02$  & 1.44$\pm0.07$  & 64 & 23  \\
156 & SIS  & 0.5  & 0.93$\pm0.03$ & 790 & 24 \\
\tableline
162 & NFW  & 0.25$\pm0.01$  & 0.71$\pm0.04$  & 52 & 23 \\
162 & Hernquist  & 0.53$\pm0.02$ & 1.93$\pm0.10$ & 78 & 23  \\
162 & SIS  & 0.5  & 0.69$\pm0.02$  & 1128 & 24 \\
\tableline
\end{tabular}
\end{center}
\end{table*}

\begin{deluxetable}{lcccc}  
\tablecolumns{5}  
\tablewidth{0pc}  
\tablecaption{Spectroscopic Data for A2199\tablenotemark{a}\label{redshifts}}  
\small
\tablehead{  
\colhead{}    RA & DEC & $cz$ & $\sigma _{cz}$ & Reference\\
\colhead{}    (J2000) & (J2000) & ($\kms$)  & ($\kms$) &\\
}
\startdata
 15 57 54.36 & +41 56 13.8 & 10387 & 125 & 04 \\ 
 15 57 58.01 & +41 47 32.6 & 10464 & 033 & 04 \\ 
 15 57 59.40 & +40 02 02.0 & 09300 & 000 & 04 \\ 
 15 58 47.35 & +41 56 17.2 & 11561 & 033 & 04 \\ 
 15 59 28.85 & +39 49 39.5 & 09296 & 045 & 04 \\ 
\enddata	     						  
\tablenotetext{a}{The complete version of this table is in the
 electronic edition of the Journal.  The printed edition contains only
 a sample.} 
\tablerefs{
(1) FAST spectra (2) Mohr et al.~2002, in prep.; (3) Mahdavi et
al.~1999; (4) Falco et al. 1999; (5) Kirshner et al.~1983: (6) Strauss
\& Huchra 1988; (7) de Vaucouleurs et al.~1991; (8) Freudling et
al.~1992; (9) Zabludoff et al.~1993; (10) Haynes et al.~1997; (11)
Hill \& Oegerle 1998.}
\end{deluxetable}

\begin{table*}[th] \footnotesize
\begin{center}
\caption{\label{xdata} \sc Archival Pointed X-ray Observations}
\begin{tabular}{lcccc}
\tableline
\tableline
\tablewidth{0pt}
System & Satellite & Instrument & Sequence & Live Time(ks) \\ 
\tableline
A2199 & ASCA & GIS/SIS &80023000 & 67 \\
	 & ROSAT & PSPC & rp800644n00 & 41 \\
A2197W &ROSAT & PSPC & rp800363n00/a01 & 13.5 \\
 &ASCA & GIS/SIS & 85068000 & 45 \\
A2197E & ROSAT & PSPC & rp800363n00/a01 & 13.5 \\
 & ASCA & GIS/SIS & 85069000 & 49 \\
NRGs385 & ROSAT & PSPC &  rp700508n00/a01 & 5.6 \\
\tableline
\end{tabular}
\end{center}
\end{table*}

\begin{table*}[th] \footnotesize
\begin{center}
\caption{\label{properties} \sc Properties of Systems Associated with A2199}
\begin{tabular}{lcccccccc}
\tableline
\tableline
\tablewidth{0pt}
System &\multicolumn{2}{c}{X-ray Coordinates} & $cz$ & $\sigma _p$ & $R_p$ &
 log$L_X$ & $N_{3\sigma_p}$ & $N_{tot}$\\ 
 & RA (J2000) & DEC (J2000) & $\kms$ & $\kms$ & $\Mpc$ &
 $h^{-2}$~ergs~s$^{-1}$ & &\\
\tableline
A2199 & 16 28 38 & 39 33 05 & 9101$\pm50$ & 796$^{+38}_{-33}$ & -- &
44.1 & 255 & 339\\
A2197W & 16 27 41 & 40 55 40 & 9144$\pm 52$ &584$^{+40}_{-33}$ & 2.2 &
42.5 &128 & 144\\
A2197E & 16 29 43 & 40 49 12 & 9100$\pm 53$& 595$^{+42}_{-34}$ & 2.0 & 42.4 & 126 & 151\\
NRGs385 & 16 17 15 & 34 55 00 & 9308$\pm86$ & 643$^{+69}_{-52}$ & 8.1
& 42.8 & 59 & 66 \\
NRGs388 & 16 23 01 & 37 55 21 & 9421$\pm74$ & 563$^{+61}_{-46}$ & 3.1
& 42.3 & 58 & 78 \\
NRGs396 & 16 36 50 & 44 13 00 & 9554$\pm128$  & 513$^{+128}_{-73}$ & 7.7 &
42.0$^{\tablenotemark{a}}$ & 16 & 29\\
NGC 6159 & 16 27 25 & 42 40 27 & 9566$\pm95$  & 344$^{+100}_{-53}$ & 4.9 &
42.3$^{\tablenotemark{b}}$ & 13 & 22\\
\tableline
\tablenotetext{a}{Detection significance = $2.7 \sigma$}\\
\tablenotetext{b}{HRI observation \citep{2000A&A...353..487T}}\\
\end{tabular}
\end{center}
\end{table*}

\begin{table*}[th] \footnotesize
\begin{center}
\caption{\label{x97} \sc Properties of X-ray Sources in A2197
field with Optical Counterparts}
\begin{tabular}{lcccccccc}
\tableline
\tableline
\tablewidth{0pt}
ID & Type &\multicolumn{2}{c}{X-ray Coordinates (J2000)} & $cz$ &
 $f_X$(0.1-2.4 keV) &  log$L_X$ \\ 
 & & RA  & DEC  & $\kms$ & $10^{-12}$ergs~cm$^{-2}$~s$^{-1}$ &
 $h^{-2}$~ergs~s$^{-1}$ \\
\tableline
A2197W & Group & 16 27 41 & 40 55 40 & 9408$\pm 77$\tablenotemark{a} &
3.13$\pm$0.10 & 42.5 \\
A2197E & Group & 16 29 43 & 40 49 12 & 8842$\pm 45$\tablenotemark{a} & 2.57$\pm$0.09 & 42.4 \\
NGC 6146 & Galaxy & 16 25 10 & 40 53 34 & 8879$\pm 25$ & 0.18$\pm$0.03 & 41.3 \\
NGC 6150\tablenotemark{b} & Galaxy & 16 25 50 & 40 29 19 & 8707$\pm 31$ & 0.04$\pm$0.01 & 40.6 \\
NGC 6175 & Galaxy & 16 29 58 & 40 37 43 & 8986$\pm 34$ & 0.60$\pm$0.04 & 41.8 \\
III Zw 078\tablenotemark{b} & Galaxy & 16 26 46 & 40 41 40 & 8838$\pm 40$ & 0.07$\pm$0.01 & 40.9 \\
Anonymous\tablenotemark{c} & Galaxy & 16 28 26 & 40 51 39 & 9976$\pm 24$ & 0.10$\pm$0.01 & 41.0 \\
2MASXi J1626069+412046 & Galaxy & 16 26 07 & 41 20 39 & 17592$\pm 21$ & 0.05$\pm$0.01 & 41.3 \\
A2187 & Cluster & 16 24 11 & 41 13 46 & 54938$\pm 54$\tablenotemark{a} & 2.37$\pm$0.08 & 44.2 \\
A2196 & Cluster & 16 27 26 & 41 29 55 & 38811$\pm 30$\tablenotemark{a} & 1.28$\pm$0.06 & 43.6 \\
FBQS J162901.3+400759 & QSO & 16 29 01 & 40 08 34 & 81545\tablenotemark{d} & 1.41$\pm$0.06 & 44.3 \\
KUV 16295+4054 & QSO & 16 31 13 & 40 48 40 & 77048\tablenotemark{e} & 1.54$\pm$0.08 & 44.5 \\
\tableline
\tablenotetext{a}{Redshift of bright galaxy near center}\\
\tablenotetext{b}{tentative identification, X-ray-optical separation $\sim 1^\prime$}\\
\tablenotetext{c}{see also \citet{1998AJ....115.1737K}}\\
\tablenotetext{d}{\citet{1995A&AS..110..469B}}\\
\tablenotetext{e}{\citet{1992AJ....104.1706C}}\\
\end{tabular}
\end{center}
\end{table*}

\begin{table*}[th] \footnotesize
\begin{center}
\caption{\label{mpfits} \sc Mass Profile Fit Parameters}
\begin{tabular}{llccccccc}
\tableline
\tableline
\tablewidth{0pt}

Method & Profile  &$q$ &$a$ & $1-\sigma$ & $M(a)$ &$1-\sigma$ & $\chi ^2$ & $\nu$ \\
& & & $\Mpc$ & $\Mpc$ & $10^{14} M_\odot$ & $10^{14} M_\odot$ & & \\
\tableline 
Standard & NFW & 10 & 0.17 & 0.13-0.21 & 0.51 & 0.45-0.58 & 1.2 & 24 \\
Standard & Hernquist & 10 & 0.50 & 0.43-0.59 & 1.75 & 1.60-1.90 & 1.1 & 24  \\
Standard & SIS & 10 & 0.5 & -- & 1.01 & 0.97-1.06 & 36.1 & 25 \\
\tableline
Standard & NFW & 25 & 0.14 & 0.11-0.19 & 0.47 & 0.40-0.55 & 0.5 & 25 \\
Standard & Hernquist & 25 & 0.47 & 0.39-0.57 & 1.68 & 1.50-1.85 & 1.2 & 25  \\
Standard & SIS & 25 & 0.5 & -- & 1.00 & 0.94-1.05 & 28.8 & 26 \\
\tableline
Standard & NFW & 50 & 0.15 & 0.10-0.23 & 0.49 & 0.36-0.65 & 0.2 & 23 \\
Standard & Hernquist & 50 & 0.46 & 0.33-0.64 & 1.65 & 1.35-2.05 & 0.4 & 23  \\
Standard & SIS & 50 & 0.5 & -- & 1.08 & 0.96-1.18 & 7.5 & 24 \\
\tableline 
$cz_{sup}=9000$ & NFW & 25 & 0.13 & 0.10-0.15 & 0.43 & 0.37-0.48 & 1.7 & 23 \\
$cz_{sup}=9000$ & Hernquist & 25 & 0.42 & 0.36-0.49 & 1.53 & 1.40-1.66 & 1.5 & 23  \\
$cz_{sup}=9000$ & SIS & 25 & 0.5 & -- & 0.97 & 0.93-1.02 & 52.0 & 24 \\
\tableline 
$cz_{sup}=9500$ & NFW & 25 & 0.30 & 0.26-0.34 & 0.79 & 0.73-0.86 & 2.1 & 61 \\
$cz_{sup}=9500$ & Hernquist & 25 & 0.94 & 0.86-1.04 & 2.80 & 2.62-2.96 & 10.2 & 61  \\
$cz_{sup}=9500$ & SIS & 25 & 0.5 & -- & 0.81 & 0.78-0.85 & 121 & 62 \\
\tableline 
Relaxed & NFW & 25 & 0.15 & 0.11-0.19 & 0.46 & 0.41-0.52 & 0.6 & 54 \\
Relaxed & Hernquist & 25 & 0.63 & 0.52-0.75 & 1.96 & 1.82-2.11 & 2.5 & 54  \\
Relaxed & SIS & 25 & 0.5 & -- & 0.59 & 0.56-0.62 & 69.0 & 55 \\
\tableline 
\end{tabular}
\end{center}
\end{table*}

\begin{table*}[th] \footnotesize
\begin{center}
\caption{\label{lxsig} \sc Predicted and Observed Velocity Dispersions}
\begin{tabular}{lccccc}
\tableline
\tableline
\tablewidth{0pt}
System & $L_X$ & $T_X$ & $\sigma _{p, L_X}$ & $\sigma _{p, T_X}$ & $\sigma _p$  \\ 
\tableline
A2199 & 44.1 & 4.5 & 892 & 810 & 796$^{+38}_{-33}$  \\
A2197W & 42.5 & 1.55 & 357 & 423 & 584$^{+40}_{-33}$   \\
A2197E & 42.4 & 0.96 & 337 & 316 & 595$^{+42}_{-34}$  \\
NRGs385 & 42.8 & -- & 424 & -- & 643$^{+69}_{-52}$   \\
NRGs388 & 42.3 & -- & 318 & -- & 563$^{+61}_{-46}$    \\
NRGs396 & 42.0 & -- & 268 & -- & 513$^{+128}_{-73}$  \\
NGC6159 & 42.3 & -- & 318 & -- & 344$^{+100}_{-53}$  \\
\tableline	                            
\end{tabular}	
\end{center}	                  
\end{table*}

\end{document}